\documentclass[12pt,a4paper,oneside]{article}

\usepackage{fourier}
\usepackage[T1]{fontenc}
\usepackage{amsfonts}
\usepackage{amssymb}
\usepackage{amsthm}
\usepackage{amsmath,graphicx,caption,mathtools}
\usepackage{relsize}
\usepackage{fullpage}
\usepackage{subcaption}
\usepackage{booktabs}
\usepackage{changepage}
\usepackage{natbib}
\usepackage[final, bookmarks, bookmarksnumbered=true, bookmarksopen=true, colorlinks=true, hyperfootnotes, linkcolor=black, anchorcolor=blue, citecolor=blue, filecolor=blue, menucolor=blue, runcolor=blue, urlcolor=black]{hyperref}
\usepackage{cleveref}
\usepackage{threeparttable}
\usepackage{tikz}
\usetikzlibrary{automata,positioning,fit, shapes.geometric,calc}
\usepackage{algorithm}
\usepackage[]{algpseudocode}
\usepackage{bbm}
\usepackage{array}
\usepackage{comment}
\usepackage{eurosym}

\newcommand\blfootnote[1]{
  \begingroup
  \renewcommand\thefootnote{}\footnote{#1}
  \addtocounter{footnote}{-1}
  \endgroup
}

\newtheorem{assumption}{Assumption}

\newtheorem{proposition}{Proposition}

\begin{document}

\title{Teacher bias or measurement error?}
\author{Thomas van Huizen\thanks{Department of Applied Economics, Utrecht University School of Economics, Utrecht University, Utrecht, The Netherlands. t.m.vanhuizen@uu.nl} \and 
Madelon Jacobs\thanks{Research Centre for Education and the Labour Market (ROA), School of Business and Economics, Maastricht University, Maastricht, The Netherlands. mce.jacobs@maastrichtuniversity.nl} \and 
Matthijs Oosterveen\thanks{Department of Economics, Lisbon School of Economics and Management, and Advance/ISEG Research, University of Lisbon, Lisbon, Portugal. oosterveen@iseg.ulisboa.pt}
\blfootnote{
We thank S{\"o}nke Matthewes, Joppe de Ree, Astrid Sands{\o}r, and Dinand Webbink for valuable comments. The paper has also benefited from the comments of participants at the EQOP meeting at the University of Oslo, the workshop on exam performance and test-taking behavior at University Carlos III of Madrid, KVS new paper sessions,  LMG Lisbon, the EALE, LEER,  LESE, and PPE conference, and the brown-bag seminars at the University of the Balearic Islands, the University of Bergen, and the Utrecht University School of Economics. 
The data used in this study are housed within the secure virtual environment at Statistics Netherlands and cannot be publicly shared due to privacy restrictions. However, all the underlying data supporting the presented results are securely accessible through Statistics Netherlands. National and international researchers can request access to the administrative data through Statistics Netherlands, and to the test score data through the Netherlands Cohort Study on Education (NCO) by contacting info@nationaalcohortonderzoek.nl.
Matthijs Oosterveen gratefully acknowledges financial support from FCT,  Fundação para a Ciência e a Tecnologia (Portugal), national funding through research grant UIDB/04521/2020 \& UID06522.
The authors have no relevant or material financial interests that relate to the research described in this paper. All omissions and errors are our own. Author names are ordered alphabetically.
}}
\date{\today \\ \vspace{0.2in}}

\maketitle
\thispagestyle{empty}

\vspace{-.3in}
\begin{abstract} 
\noindent {\normalsize 
Subjective teacher evaluations play a key role in shaping students' educational trajectories. Previous studies have shown that students of low socioeconomic status (SES) receive worse subjective evaluations than their high SES peers, even when they score similarly on objective standardized tests. This is often interpreted as evidence of teacher bias. Measurement error in test scores challenges this interpretation. We discuss how both classical and non-classical measurement error in test scores generate a biased coefficient of the conditional SES gap, and consider three empirical strategies to address this bias. Using administrative data from the Netherlands, where secondary school track recommendations are pivotal teacher judgments, we find that measurement error explains 35 to 43\% of the conditional SES gap in track recommendations.
} 


\end{abstract}

\linespread{1.50}
\normalsize

\newpage
\setcounter{page}{1}

\section{Introduction}

Teacher bias may be central to inequalities in educational attainment and later life outcomes. To estimate the extent of such bias, researchers typically compare student assessments in which teachers exercise discretion, such as teacher-assigned grades or track recommendations, to more objective indicators of student ability, such as standardized test scores \citep{alesina2024revealing,falk2023mentoring}. However, the presence of systematic gaps in subjective teacher evaluations between groups, conditional on objective ability measures, is not necessarily evidence of teacher bias. This approach faces two main concerns: omitted variables and measurement error in the ability measure. Studies often address the first concern by controlling for additional factors \citep{burgess2013test}. In contrast, the second concern is typically overlooked or addressed under strict assumptions that may not hold in practice, a shortcoming shared by the broader literature on discrimination.

This paper focuses on the second concern: measurement error in the objective measure of ability. In classical testing theory, test scores are typically based on the number of correct answers and provide an unbiased but noisy estimate of ability. The resulting ``classical'' measurement error in the test score is considered purely random and hence uncorrelated with ability. In modern item response theory (IRT), statistical models produce test scores that may be biased estimates of ability but are unbiased as predictors. The resulting ``non-classical'' measurement error may correlate with ability. For instance, in case the test score is estimated with a posterior mean score, students with a low and high signal are drawn towards the population mean, such that the error correlates negatively with ability, yet is uncorrelated with the test score. 

To examine the implications, consider a regression in which a teacher’s subjective evaluation of a student is the dependent variable, and the independent variables include a group characteristic, such as family socioeconomic status (SES), and a test score as a measure of student ability. In case of classical measurement error, the test score coefficient is biased towards zero (the well-known attenuation bias). Moreover, if student ability is associated with SES and matters for teacher evaluations, this bias also contaminates the coefficient of the conditional SES gap. In particular, when ability is positively associated with both SES and teacher evaluations, the conditional SES gap is overestimated. In case of non-classical measurement error, the attenuation bias in the test score coefficient may be partially or fully eliminated, though this leads to a biased coefficient of the association between ability and SES. This bias then also contaminates the estimate for the conditional SES gap.

We first show how both classical and certain forms of non-classical measurement error, particularly relevant to modern test scores, generate identical contamination bias in the coefficient measuring the conditional gap between groups. We then discuss three empirical strategies to address measurement error. First, we consider an instrumental variable (IV) approach, a well-established method in econometrics to correct for measurement error \citep{hausman2001mismeasured}. Second, we discuss a method in the spirit of errors-in-variables (EIV) models, an approach that is common in psychometrics \citep{fuller2009measurement}. The additional parameter required for the EIV strategy, compared to OLS, is the reliability ratio of the test that is used to measure ability. We argue that, under weaker assumptions than the IV strategy, the IV first stage can be used to identify this ratio. Third, we combine the EIV strategy with a novel approach to identify the reliability ratio under yet weaker assumptions. 

We apply these three strategies to estimate conditional gaps in teacher track recommendations in the Netherlands, focusing on differences by family SES. The Dutch tracking regime provides a compelling setting for this analysis. While nearly all OECD countries have some form of educational tracking, the Netherlands assigns students to secondary-school tracks at a relatively young age of twelve. Assignment is based on binding track recommendations from the primary school teacher \citep{wvo2014}. These recommendations may be highly consequential, as roughly 70\% of students does not experience track mobility four years into secondary education \citep{deree2023quality}. Track placement determines the level of the curriculum and peer group abilities, and may affect student achievement \citep{matthewes2021better}. As only the highest tracks provide access to university, track placement may also have long-term consequences for attainment and income \citep{dustmann2017long,borghans2019long}. Moreover, negatively biased recommendations signal low teacher expectations, which can in turn harm student outcomes \citep{carlana2019implicit,papageorge2020teacher}. 

Given the consequential nature of these recommendations, potential teacher bias has been a major policy concern in the Netherlands. This concern gained momentum in 2016, when a report by the Dutch \citet{inspectorate2016} demonstrated that students from low SES families receive significantly lower track recommendations, conditional on scores from a standardized end-of-primary school test. This finding fueled an ongoing public debate on teachers as gatekeepers of opportunities and contributed to a national policy reform, effective from 2023-24. The reform reduces the role of teachers in track placement, and increases the role of the end-of-primary school test. Similar findings in academic studies reinforce concerns about unequal opportunities in education, both in the Netherlands \citep{zumbuehl2025can} and elsewhere \citep{carlana2022implicit,falk2023mentoring}.

Using new administrative data, we first replicate previous findings from the Netherlands, and find that children with lower family SES receive systematically lower track recommendations, conditional on standardized test scores. We discuss that our test scores are derived from an IRT model, with ability parameters estimated using weighted maximum likelihood \citep{sanders1993psychometrie}. This method is known to produce largely unbiased estimates of student ability \citep{warm1989weighted}, and the measurement error is considered classical. We address concerns about measurement error using three strategies. Our novel strategy combines the EIV method with an unbiased estimate for the test's reliability ratio under weak assumptions, based on students’ standardized test scores across all primary school grades. Across the three approaches, we find that measurement error can explain between 35 and 43\% of the conditional SES gap. 

Our paper contributes to the broad literature on measurement error \citep{schennach2016recent}. In particular, we focus on contamination bias due to measurement error. As \citet{modalsli2022spillover} point out,
``\textit{[a]lthough the notion of bias in one coefficient arising from error in another regressor is a well-known econometric result, it is seldom addressed in practice with empirical studies.}''
\citet{modalsli2022spillover} examine the implications of classical measurement error in multigenerational income regressions and use an IV approach to mitigate the resulting bias. Similarly, \citet{gillen2019experimenting} use IV to show that the gender gap in competitiveness is well explained by risk attitudes and overconfidence once classical measurement error is addressed. Our study extends this literature by applying three strategies to address contamination bias and by incorporating both classical and non-classical measurement error. The latter is pervasive in education, where test scores are often derived from IRT models \citep{jacob2016measurement}. 
It is also common in the labor market, as many individuals misreport hours worked, leading to biased estimates of wage discrimination and inequality \citep{borjas2024}. More generally, it arises in survey data when respondents base their answers on incomplete information \citep{hyslop2001bias}

We also contribute to the literature on teacher bias by providing new evidence on the conditional SES gap in track recommendations. Teacher bias in track recommendations has been studied in Germany \citep{falk2023mentoring}, Italy \citep{carlana2022implicit}, and the Netherlands \citep{zumbuehl2025can}. Previous studies have also examined teacher bias in grading by gender \citep{cornwell2013noncognitive,lavy2018origins,terrier2020boys}, ethnicity \citep{burgess2013test}, and migration background \citep{alesina2024revealing}. Consistent with the observation above, contamination bias due to measurement error has been largely ignored in this literature. Three notable exceptions are \citet{botelho2015racial}, \cite{ferman2022assessing}, and \cite{zhu2024}, who study teacher bias in grading in Brazil and the US. All three studies find that the conditional gaps are substantially reduced or even reversed when addressing measurement error in test scores. However, as these studies assume classical measurement error and exclusively apply an IV strategy, it remains unclear whether the results hold under non-classical error and are robust to alternative strategies relying on weaker assumptions.

\section{Educational tracking in the Netherlands}\label{sec:Setting}

In the Netherlands, the transition to tracked secondary education takes place around the age of twelve when students leave the sixth and final grade of primary education. The secondary school system consists of three main tracks: (i) a university track (\textit{vwo}), which has a duration of six years and gives students access to university education, (ii) a college track (\textit{havo}), which has a duration of five years and prepares students for college education, and (iii) a vocational track (\textit{vmbo}), which has a duration of four years and serves as a preparation for vocational education. The vocational track consists of four sub-tracks, which can be ordered from more practice-oriented to more theory-oriented. 

The allocation of students to tracks is based on the recommendation provided by the sixth-grade primary school teacher. The recommendation process consists of two main steps. First, by March of the sixth grade the teacher provides an initial track recommendation. Then, in April or May, each student takes a standardized end-of-primary education test, which the teacher may use to consider an upward revision \citep{wpo2014}. Hence, the final track recommendation may be higher (but not lower) than the initial one. The teacher track recommendation is binding in the sense that, as a rule, secondary schools cannot place students in a higher track than indicated by the recommendation \citep{wvo2014}.

We use the initial, instead of the final, track recommendation as our measure for the teacher's subjective evaluation. This choice reflects that the initial recommendation is often the most influential, as many secondary schools begin allocating students to first-year classes as early as March, relying on these initial recommendations. Secondary schools may face challenges when changing their allocation by the time the final track recommendation is provided in May \citep{inspectorate2019}. Moreover, in practice, teachers upgrade only 1 in 10 students, meaning the initial and final track recommendations are identical for 9 out of 10 students \citep{deree2023quality}.

Teachers are expected to make holistic track recommendations by considering both objective standardized test results and their subjective judgments on factors such as motivation, attitude towards school, and classroom behavior. While there are no strict national rules describing how to combine these types of information, general guidelines from the \cite{ministry2022} suggest that standardized test results provide a ``good starting point''. A recent survey among 400 primary school teachers shows that 85\% of the respondents attach ``considerable'' to ``a lot of'' value to standardized test results when formulating their recommendations \citep{Duo2023}. 

\subsection{Standardized test scores}\label{sec:Testscores}

All Dutch primary school students, from grade one to six, are expected to take two standardized tests per year. The first test is administered in the middle of the school year around February and the second at the end of the school year in June. We will refer to these tests as the midterm and end term test, respectively. We use the fifth-grade end term test as the main measure of student ability to estimate the conditional SES gap in track recommendations. This test can be considered the last score that teachers observe before making the initial recommendation. Alternatively, one could use the end-of-primary school test score (the sixth-grade end term test) as the ability measure and the final track recommendation as the teacher's subjective evaluation. This alternative approach is problematic since the final track recommendation can only be adjusted upwards based on performance on the end-of-primary school test. Hence, this test is low (high) stakes for students who are (un)satisfied with the initial recommendation.

Primary schools have to select one of the standardized test systems that are approved by the Ministry of Education. Most schools opt for the test system provided by \textit{Cito}, which has a market share of roughly 80\% in the period we study. The biannual \textit{Cito} tests focus on two domains: mathematics and reading. The tests consist of approximately 70 to 100 items, include both multiple-choice and open-ended questions, and are scored by the teacher using standardized answer keys. The item scores are entered into the \textit{Cito} software and used to generate students' test scores through an item response theory (IRT) model. 

An IRT model specifies the probability that a student answers a test item correctly as a function of latent student ability and item parameters. The IRT model used by \textit{Cito} is the Birnbaum model \citep{Cito2017}, which includes two item parameters: difficulty and discrimination. The item difficulty parameter reflects the student ability level at which there is a 50\% probability of answering the item correctly. The discrimination parameter reflects how sharply the item distinguishes between students of different ability levels. These parameters may be estimated for a selection of items to form a so-called calibrated item bank, which can be used to create multiple tests of varying difficulty that aim to measure student ability on a common scale. This enables, for instance, observing the development of student ability from grade 1 to 6, one of the main goals of standardized testing in the Netherlands. 

\textit{Cito} constructs its calibrated item bank through national norming studies. First, sample tests are administered to representative student populations. Second, the item parameters are estimated: the discrimination parameters are held fixed while the difficulty parameters are estimated using conditional maximum likelihood. Both parameters are then iteratively adjusted to improve model fit. The resulting calibrated item bank enables the construction of the biannual tests for various grade levels. Although more difficult items are used in tests for higher grades, student ability can supposedly be measured on a common scale.

Latent student ability is estimated from these biannual tests as a ``fixed effect'' using weighted maximum likelihood \citep{sanders1993psychometrie}. In this approach, the two item parameters are held fixed at their calibrated values, and the likelihood function is weighted by the test information function. This function indicates how much information the test provides about student ability, with items of difficulty close to the student’s ability offering more information. \cite{warm1989weighted} shows that this procedure yields (nearly) unbiased estimates of student ability. The estimated ability scores obtained through this procedure are the test scores we use in our analysis. Unbiased, yet noisy, estimates of ability implies that the measurement error in the biannual \textit{Cito} test scores is assumed to be classical.

\section{Data}\label{sec:Data}

We use proprietary administrative data from Statistics Netherlands. The data on the biannual standardized tests is gathered by the Netherlands Cohort Study on Education (NCO) and subsequently made available to Statistics Netherlands. The NCO collected the test score data only from schools using the tests developped by \textit{Cito}. Since 2018–19, data were obtained by individually contacting primary schools, each of which had to grant permission. As a result, the test score data covers about 50\% of all primary schools in the Netherlands. See \cite{haelermans2020using} for details on the NCO data collection. This test score data can be linked to standard administrative data using an anonymized personal ID, allowing us to also observe teacher track recommendations, parental income and education, and several other student background characteristics.

The test score history, from grade one to six, is available for the students who were enrolled in sixth grade from 2018-19 onward. Our sample includes four cohorts: students who are in sixth grade in the school years 2018-19, 2019-20, 2020-21, and 2021-22. The standard administrative data covers 686,309 students across these four cohorts, with almost no missing values. The biannual test score data includes 318,630 students, although not all students took both standardized tests in each grade. After merging these two datasets, we observe 276,841 students in 5,802 schools. After dropping students with one or both grade 5 test scores missing, as well as those with extreme household parental income values (discussed below), our main estimation sample consists of 148,019 students in 3,295 schools.

\subsection{Descriptive statistics}

\begin{table}
\small
\linespread{1.00}
\begin{center}
\caption{Descriptive statistics}
\label{tbl:descrstats}
\begin{tabular}{lcccc|ccc}
\toprule \toprule
&\multicolumn{4}{c}{Estimation sample}&\multicolumn{3}{c}{Samples w/ missings}\\
[-.75em]
&\multicolumn{4}{c}{}&\multicolumn{3}{c}{Admin \, Test score \, Merged}\\ 
&\multicolumn{1}{c}{(1)}&\multicolumn{1}{c}{(2)}&\multicolumn{1}{c}{(3)}&\multicolumn{1}{c}{(4)}&\multicolumn{1}{c}{(5)}&\multicolumn{1}{c}{(6)}&\multicolumn{1}{c}{(7)}\\
[-.50em]
& Mean & SD  & Min & Max & \multicolumn{3}{c}{Mean} \\
\midrule
&\multicolumn{7}{c}{Panel A: Teacher recommendation} \\
$\mathbbm{1}$(recom. $\geq$ upper voc. tr.)&       0.770&       0.421&       0.000&       1.000 &0.754&&0.758 \\
$\mathbbm{1}$(recom. $\geq$ college tr.)&       0.497&       0.500&       0.000&       1.000 &0.478&&0.489\\
$\mathbbm{1}$(recom. $\geq$ university tr.)&       0.211&       0.408&       0.000&       1.000 &0.197&&0.209\\
[.25em]
&\multicolumn{7}{c}{Panel B: SES measures} \\
SES income          &       6.439&       3.772&      $<$-0.410 &      $>$55.500 &6.394&&6.472\\
$\mathbbm{1}$(SES inc. $\geq$ median)&       0.517&       0.500&       0.000&       1.000 &0.500&&0.506\\
SES years of schooling&      15.975&       2.334&       8.000&      18.000 &16.004&&15.945\\
$\mathbbm{1}$(SES edu. $\geq$ college)&       0.559&       0.497&       0.000&       1.000 &0.618&&0.553\\
[.25em]
&\multicolumn{7}{c}{Panel C: Test scores} \\
Reading midterm grade 5     &     191.618&      26.766&      $<$81.000&     $>$340.000 &&191.614&192.536\\
Reading end term grade 5    &     197.279&      26.983&       $< $84.000&      $> $354.000 &&195.374&196.367\\
Math midterm grade 5     &     254.142&      27.142&       $< $118.000&      $> $379.000 &&253.982&254.564\\
Math end term grade 5    &     261.959&      26.085&       $< $118.000&      $>$394.000 &&262.583&263.179\\
Avg. midterm grade 5     &       0.005&       0.900&      -5.177&       5.077 &&&\\
Avg. end term grade 5    &       0.004&       0.905&      -5.438&       5.130 &&&\\
[.25em]
&\multicolumn{7}{c}{Panel D: Control variables} \\
$\mathbbm{1}$(female)&       0.499&       0.500&       0.000&       1.000 &0.503&&0.503\\
$\mathbbm{1}$(Western imm.)&       0.076&       0.265&       0.000&       1.000 &0.077&&0.077\\
$\mathbbm{1}$(non-Western imm.)&       0.200&       0.400&       0.000&       1.000 &0.178&&0.198\\
Age                 &      11.590&       0.647&        $< $9.250&      $>$14.160 &11.704&&11.660\\
[.25em]
Observations        &    \multicolumn{4}{c}{148019}      &686309 &318630&  276841  \\
\bottomrule \bottomrule
\end{tabular}
\begin{tablenotes}
\footnotesize
\item[] Notes: this table shows the descriptive statistics across four samples. Income is measured in \texteuro 10,000.
We translated the schooling levels in the Netherlands to years of schooling as follows: primary school $=$ 8, mbo 1 $=$ 13 , havo or vwo $=$ 14, mbo 2 or 3 $=$ 15, mbo 4 $=$ 16, college or university bachelor $=$ 17, and university master $=$ 18 years.
Statistics Netherlands makes a distinction between individuals with a Western and non-Western migration background. Non-Western immigrants have a background from all countries in Africa, Latin America or Asia (including Turkey, excluding Indonesia and Japan), and Western immigrants from the remaining countries.
Statistics Netherlands prevents reporting minima and maxima that are based upon less than 10 observations. Therefore, the minimum (maximum) may contain the smallest (largest) number, such that at least ten observations have values lower (higher) than that number. 
\end{tablenotes}
\end{center}
\end{table}

The first four columns in \Cref{tbl:descrstats} present the descriptive statistics for our estimation sample. Panel A focuses on the measures used for the subjective teacher evaluation. Our primary measure is a dummy variable that equals one if the initial track recommendation is at least equal to the college track. College or university track enrollment is highly policy relevant, since it implies the student is tracked towards higher education. Nearly half (49.7\%) of the students receive a track recommendation equal to, or higher than, the college track.

Panel B presents the statistics for the different measures of SES. Our baseline SES measure is gross annual household parental income, which averages \euro64,390. We dropped observations with a yearly parental income above (below) the 99.9th (0.1st) percentile. In our analysis below we standardize parental income per cohort. For robustness checks, we use three alternative SES measures: parental years of schooling of the highest educated parent, a binary income indicator denoting whether parental income exceeds the student's cohort median, and a binary education indicator reflecting whether the highest educated parent holds at least a college degree.

Panel C shows the statistics for the biannual \textit{Cito} tests in fifth grade, per domain and the average across domains. The domain-specific scores are the raw scores estimated on a common scale by the IRT model. The average score is calculated by first standardizing the domain-specific scores per cohort, and subsequently taking the mean across domains. \Cref{tbl:descrstats_tests} in Appendix D provides the descriptive statistics for the raw domain-specific scores on the biannual tests in grades two through four. We exclude grade 1 data since a relatively high fraction of the test scores for that grade are missing. Panel D presents the statistics for the variables used as controls in the robustness checks.

We also compare the statistics of the students in our estimation sample with those from three larger samples representing the different steps of our selection process. Columns (5) to (7) correspond to the samples from the standard administrative data covering the full population, the test score data, and the merged dataset, respectively. The statistics in these columns are based on different numbers of observations due to missing values in some variables, particularly in the grade 5 test scores of columns (6) and (7). The comparison indicates that our estimation sample closely resembles the student population in the Netherlands.

\section{Methodology}\label{sec:Model}

Let $y_t$ be a variable that measures a teacher's subjective evaluation of a student in time period $t$, $SES$ a variable that measures the socioeconomic status of the student, and $s_t$ a variable that reflects student ability in time period $t$. The long regression equation equals,
\begin{align} \label{eqn:long}
    y_{t}= \beta SES + \gamma s_{t}+e_{t}.
\end{align}

Throughout the paper we suppress the subscript for student $i$ and the intercept in regression equations. Scholars and policy makers are generally interested in the parameter $\beta$. That is, they are interested in whether subjective teacher evaluations differ systematically by SES, or another group characteristic such as gender or ethnicity, conditional on student ability. If individuals with similar student ability experience similar causal effects of the subjective evaluation on outcomes of interest, then $\beta$ can also be interpreted as a measure of principal fairness, as recently introduced by \cite{imai2023principal}.

\subsection{Classical versus non-classical measurement error}

We cannot estimate \eqref{eqn:long} as we do not observe ability. Instead, students produce an unbiased but noisy signal of ability by taking a test, denoted by $\widetilde{s}_{t}$. The signal equals ability plus measurement error, $\widetilde{s_t}=s_t+m_t$, where $m_t$ is the signal's measurement error. An unbiased signal, meaning one that equals ability in expectation, implies that the error is classical.
\begin{assumption}\label{ass:classical}
Classical measurement error in the signal
\renewcommand{\theenumi}{\alph{enumi}}
\begin{enumerate}
    \item[] (Zero mean and cov) $\mathbb{E}[m_{t}]=0$ and $\mathrm{Cov}[m_{t},x]=0$ $\forall x \neq \widetilde{s}_t, t$. 
\end{enumerate} 
\end{assumption}
This classical error reflects, for instance, that students may guess answers, are more or less familiar with the specific selection of test items, or feel (un)well on the day of the test. According to classical testing theory, measures such as the fraction of correctly answered items follow \Cref{ass:classical} and are unbiased but noisy estimates of ability.

In modern assessment systems, test scores are rarely that simple. Instead, they are often the product of statistical IRT models designed to extract more information from the pattern of responses. For instance, \cite{jacob2016measurement} discuss that IRT test scores in many assessment systems in the US, such as the ECLS, the NELS:88, the ELS, and the HSLS are constructed from posterior means. Let $s^m_t$ denote the test score, which is then computed by 
\begin{align} \label{eqn:posterior}
   s_{t}^m=\mathbb{E}[s_t|\widetilde{s}_t] &=\int s_t f(s_t \mid \widetilde{s}_t) d s_t \\ \label{eqn:shrinking} 
   &\simeq  (1-\widetilde{\lambda} )\mathbb{E}[\widetilde{s}_t]+\widetilde{\lambda} \widetilde{s}_t, \quad \forall t.
\end{align} 
Students are treated as a random sample from a population of ability values $s_t$, characterized by the posterior density function $f(s_t \mid \widetilde{s}_t)$ from the IRT model. The posterior mean score is the mean of that population. Hence, instead of using the signal as the test score directly, the signal is used as a predictor for ability to generate a test score. This prediction is not perfect, so we can similarly write that the test score equals ability plus the score's measurement error, $s^m_{t}=s_{t}+{m}^{m}_{t}$. 

Posterior mean scores can be interpreted as Empirical Bayes estimates that, roughly, shrink the student’s (unbiased) maximum likelihood score toward the population mean in proportion to the noisiness of the maximum likelihood score \citep{jacob2016measurement}. Building on this interpretation, we approximate the posterior mean in \eqref{eqn:posterior} by the \cite{kelley1947fundamentals} estimator in \eqref{eqn:shrinking}, where $\widetilde{\lambda}$ is the shrinkage factor. With $\widetilde{\lambda}  \in [0,1)$, students with low (high) ability signal are pushed up (down) towards the population mean. As a result, the measurement error of the test score $m_t^m$ correlates negatively with ability. This feature of the non-classical error in \eqref{eqn:shrinking} is similar to the error generated by any posterior mean in \eqref{eqn:posterior}.

We will analyze contamination bias in $\beta$ when replacing ability in \eqref{eqn:long} by test scores measured through \eqref{eqn:shrinking}. Our results hold (approximately) for two types of test scores. The first are test scores estimated classically or through an IRT model using (weighted) maximum likelihood under a “fixed effect” approach. Both provide an unbiased estimate of ability. This corresponds to \eqref{eqn:shrinking} with $\widetilde{\lambda}=1$: the test score is equal to the signal, $s^m_t=\widetilde{s}_t$, and measurement error is classical, $m^{m}_t=m_t$.  The second are test scores estimated from an IRT model under a “random-effects” approach, where the scores are posterior means. Equation \eqref{eqn:shrinking} holds exactly only for the Kelley estimator, which corresponds to normality for both the prior and likelihood distributions. Generally, it is an approximation, and depending on the choice of these two distributions, the posterior mean in \eqref{eqn:posterior} may even lack a closed form. Because the Kelley estimator corresponds to an OLS regression of $s_t$ on $\widetilde{s}_t$, by the properties of OLS it provides the best linear approximation to the (possibly nonlinear) posterior mean. 

Our results speak less to settings where test scores are reported as posterior medians or modes, where multiple plausible values are drawn from a student’s posterior distribution, or where the prior distribution on ability incorporates student characteristics. The latter approach is used in the major US and international assessments NAEP and PISA. An alternative strategy is to jointly estimate the research model in \eqref{eqn:long} alongside the IRT model, treating the latter as a direct model of measurement \citep{junker2012use}. This requires access to item-level response data, which is typically unavailable. IRT models often produce noisier test scores at the tails of the distribution, reflecting the limited information available to these parametric methods when student ability does not align well with item difficulty. Our approach allows for the potential difference between so-called local and global reliability.

Non-classical measurement generated by \eqref{eqn:shrinking} with $\widetilde{\lambda} \in [0,1)$ is also considered by \cite{hyslop2001bias} in survey data. They consider a setting in which respondents are asked questions like ``what is the value of $s_t$?'' given an information set that consists of the unbiased but noisy signal $\widetilde{s}_t$. Respondents may passively report the ``flawed'' but unbiased value ($\widetilde{s}_t$), or may actively seek to provide an ``optimal'' response ($s^m_t$) given their information set. An argument for the latter is that respondents are likely aware of the lack of precision in the signal. Our results on the contamination bias below may thus also be relevant for the analysis of survey data more generally.

\subsection{Contamination bias}

\noindent The medium regression equation replaces $s_t$ with $s^m_t$,
\begin{align} \label{eqn:med}
    y_{t}= \beta^m SES + \gamma^m s_{t}^m+e_{t}^m.
\end{align}
To analyze what $\beta^m$ and $\gamma^m$ identify, we also introduce the balancing regression equation,
\begin{align}\label{eqn:balancing}
 \begin{pmatrix} s_t \\ s_{t}^m  \end{pmatrix}= \begin{pmatrix} s_t \\ \widetilde{\lambda} (s_t+m_t)  \end{pmatrix}=\begin{pmatrix} \delta \\ \delta^{m}  \end{pmatrix} SES +  \begin{pmatrix} u_{t} \\ u^m_{t} \end{pmatrix}.
\end{align} 
Note that $(1-\widetilde{\lambda} )\mathbb{E}[\widetilde{s}_t]$ is fixed and gets soaked up by the intercept of the regression. The following proposition formalizes what OLS identifies.
\begin{proposition}\label{prop:ols}
Under \Cref{ass:classical} and \Cref{eqn:shrinking}, it holds that: 
    \begin{align} 
    \label{eqn:deltam_ass}
    \delta^m=&\delta \widetilde{\lambda}, \\\label{eqn:gammam_ass}
    \gamma^m=&\gamma \Bigg(\frac{\lambda}{\widetilde{\lambda}}\Bigg), \\  
    \label{eqn:betam_ass}
    \beta^{m}=&\beta+\gamma\delta\big(1-\lambda\big).
    \end{align}    
With $\lambda=\Bigg(\frac{\sigma_{u_{t}}^2}{\sigma_{u_{t}}^2+\sigma_{m_{t}}^2} \Bigg)$, where $\sigma_{\bullet}^2$ denotes the variance of the variable in the subscript.
\end{proposition}

All proofs are provided in Appendix A to C. First consider \eqref{eqn:deltam_ass} and \eqref{eqn:gammam_ass} under classical measurement error, where $\widetilde{\lambda}=1$. Classical error on the left-hand side does not introduce bias in the estimate of $\delta^m$. However, as shown in \eqref{eqn:gammam_ass}, the estimate of $\gamma^m$ is subject to the well-known attenuation bias. The multivariate reliability (or signal-to-total variance) ratio of the test, denoted by $\lambda$, decreases from one toward zero as the measurement error variance increases, leading to attenuation bias in $\gamma^m$. Now consider the case of non-classical measurement error, where $\widetilde{\lambda} \in [0,1)$. \Cref{eqn:gammam_ass} shows that the attenuation bias in $\gamma^m$ may be smaller and disappears completely when the shrinkage factor equals the reliability ratio, $\widetilde{\lambda} = \lambda$. However, \eqref{eqn:deltam_ass} shows that non-classical error on the left-hand side results in attenuation of $\delta^m$.

\Cref{eqn:betam_ass} clarifies that both types of measurement error contaminate the estimate of $\beta^m$. Under classical measurement error, attenuation of $\gamma^m$ spills over to $\beta^m$. Under non-classical measurement error, the attenuation bias in $\gamma^m$ may be corrected, but at the expense of attenuation in $\delta^m$, which again spills over to $\beta^m$. Importantly, the contamination is identical for both types of measurement error, as the bias term involves the product of $\gamma^m$ and $\delta^m$, where the former is divided and the latter is multiplied by $\widetilde{\lambda}$. Although this may seem intuitive, to our knowledge this result has not been reported in the literature. In particular, \citet{hyslop2001bias} analyze \eqref{eqn:deltam_ass} and \eqref{eqn:gammam_ass} under both classical and non-classical measurement error, while \citet{pei2019poorly,gillen2019experimenting,ferman2022assessing} examine \eqref{eqn:betam_ass} in the case of classical measurement error only. Note that the contamination bias grows larger as $\gamma$ and $\delta$ increase and $\lambda$ decreases.
 
\subsection{The IV strategy}

The address measurement error, the instrumental variables (IV) strategy aims to use a second objective measure of ability in period $t$ as an instrument for the first measure. In practice, a lagged test score from period $t-1$ is often used instead \citep{botelho2015racial,ferman2022assessing,zhu2024}. Similar to above, students produce an unbiased but noisy signal of ability by taking a test in $t-1$, where $\widetilde{s}_{t-1} = s_{t-1}+m_{t-1}$. The corresponding test score satisfies \eqref{eqn:shrinking}, where $s_{t-1}^m=\mathbb{E}[s_{t-1}|\widetilde{s}_{t-1}]$ and $s^m_{t-1}=s_{t-1}+m^m_{t-1}$. We do not index the shrinking parameter $\widetilde{\lambda}$ by time, thereby assuming it remains constant across tests. Our main results hold without this simplification, though at the expense of additional notational clutter and reduced intuition. Using $s_{t-1}^m$ as an instrument for $s_{t}^m$ generates the following first and second stage regression equations respectively,
\begin{align}
    \label{eqn:fs}
    s_{t}^m &= \kappa SES + \pi s_{t-1}^m+e_t^{fs},\\
    \label{eqn:ss}
    y_t &= \beta^{iv} SES + \gamma^{iv} \widehat{s}_{t}^m +e_{t}^{iv}.
\end{align}

To show how this IV strategy can address measurement error, it is useful to also introduce the balancing regression for $t-1$,
\begin{align}\label{eqn:balancing_tmin1}
 \begin{pmatrix} s_{t-1} \\ s_{t-1}^m \end{pmatrix}= \begin{pmatrix} s_{t-1} \\ \widetilde{\lambda} (s_{t-1}+m_{t-1}) \end{pmatrix}=\begin{pmatrix} \delta_{t-1} \\ \delta_{t-1}^m   \end{pmatrix} SES +  \begin{pmatrix}  u_{t-1}  \\ u^m_{t-1} \end{pmatrix},
\end{align} 
and the variable $\Delta_{s_{t}}=s_t-s_{t-1}$, which captures the change in unobserved ability between time period $t$ and $t-1$. We describe two additional assumptions that place increasing restrictions on $\Delta_{s_{t}}$. 
\begin{assumption}\label{ass:fs}
The covariance between $\Delta_{s_t}$, $SES$ and $s_{t}$ is zero 
\renewcommand{\theenumi}{\alph{enumi}}
\begin{enumerate}
    \item (SES) $\mathrm{Cov}[\Delta{s_{t}},SES]=0$.
    \item (Ability) $\mathrm{Cov}[\Delta_{s_{t}},s_{t}]=0$.
    \end{enumerate} 
\end{assumption}
\noindent \Cref{ass:fs}(a) requires the change in ability to be the same, on average, across low and high SES students. \Cref{ass:fs}(b) requires this for low and high ability students in time $t$. 
\begin{assumption}\label{ass:rf}
Constant $\Delta_{s_t}$
\renewcommand{\theenumi}{\alph{enumi}}
\begin{enumerate}
    \item[] (Constant) $\Delta{s_{t}}$ is the same for each student.
    \end{enumerate} 
\end{assumption}
\noindent \Cref{ass:rf} requires that the change in ability is the same for each student, which is stronger than \Cref{ass:fs}. The following proposition summarizes what the IV strategy identifies while combining \Cref{ass:classical} with either \ref{ass:fs} or \ref{ass:rf}.
\begin{proposition}\label{prop:iv}
\indent (i) Under \Cref{ass:classical} and \ref{ass:fs}, and \Cref{eqn:shrinking}, it holds that: 
    \begin{align}
       \pi'=&\lambda, \\
         \gamma^{iv}=&\gamma \Bigg(\frac{1}{\widetilde{\lambda}}\Bigg) B_{\gamma}, \\
        \beta^{iv}=& \beta+\gamma\delta\bigg(1-B_{\gamma}\bigg).    
    \end{align} 
With $\pi'=\pi \Bigg( \frac{\sigma_{u_{t}}^2+ \sigma_{\Delta_{s_t}}^2+\sigma_{m_{t-1}}^2}{\sigma_{u_{t}}^2+\sigma_{m_{t}}^2}\Bigg)$ and $B_{\gamma}= 1 -  \Bigg(\frac{\mathrm{Cov}[e_{t},\Delta_{s_t}]}{\mathrm{Cov}[y_{t},u_{t}]}\Bigg)$.  

\vspace*{.2in}
\indent (ii) Under \Cref{ass:classical} and \ref{ass:rf}, and \Cref{eqn:shrinking}, it holds that: 
    \begin{align}
        \pi'=&\lambda, \\
        \gamma^{iv}=&\gamma \Bigg(\frac{1}{\widetilde{\lambda}}\Bigg), \\
        \label{eqn:betaiv_ass3}
        \beta^{iv}=&\beta.          
    \end{align}
With $\pi'=\pi \Bigg( \frac{\sigma_{u_{t}}^2+\sigma_{m_{t-1}}^2}{\sigma_{u_{t}}^2+\sigma_{m_{t}}^2}\Bigg)$.  
\end{proposition}

Under classical measurement error with $\widetilde{\lambda}=1$, \Cref{prop:iv} formalizes that the IV strategy yields consistent estimates under \Cref{ass:rf}. This result is not new. For instance, \cite{gillen2019experimenting} use IV to address contamination bias from classical error in \eqref{eqn:betaiv_ass3} and review similar applications in the literature. Intuitively, a constant $\Delta{s_{t}}$ implies that the test score in period $t-1$ is truly a second objective measure of ability in period $t$. The first stage regression of $s_{t}^m$ upon $s_{t-1}^m$ measures to which extent this relationship is diluted due to measurement error, so that it reveals the reliability ratio $\lambda$.
Furthermore, the reduced form equation is similar to the medium regression equation \eqref{eqn:med} while replacing $s_{t}^m$ with $s_{t-1}^m$. With a constant $\Delta{s_{t}}$ the reduced form coefficient reveals the estimate $\gamma^m=\gamma\lambda$, which was attenuated towards zero by the reliability ratio. Dividing the reduced form by the first stage corrects for this attenuation. Since $\gamma^{iv}=\gamma$ and $\delta^m=\delta$, we also have that $\beta^{iv}=\beta$.

In the presence of non-classical error with $\widetilde{\lambda}\in[0,1)$, the IV strategy also addresses contamination bias under \Cref{ass:rf}, though for a different reason. To illustrate, suppose the shrinkage factor equals the reliability ratio, such that $\widetilde{\lambda}=\lambda$. In this case, the OLS coefficient on test scores $\gamma^m$ does not suffer from attenuation bias. However, the IV strategy still divides by the first stage, which means that the IV coefficient $\gamma^{iv}$ is overestimated by a factor equal to the inverse of the reliability ratio. As noted by \cite{hyslop2001bias}, IV has a bias away from zero in the presence of non-classical measurement error.
However, this upward bias in $\gamma^{iv}$ is exactly what is needed to correct for the contamination bias to $\beta^{iv}$, which arises from attenuation in the estimate of $\delta^m$ from the balancing regression. Since the contamination bias term involves the product of the test score and balancing coefficients, the upward bias in the former offsets the downward bias in the latter. As a result, the IV strategy also effectively addresses the contamination bias in the presence of non-classical measurement error.

\subsection{The EIV strategy}

Based on the results of \Cref{prop:ols}, the errors-in-variables (EIV) strategy directly formulates expressions for the parameters of interest,
\begin{align}\label{eqn:gamma_est}
    \gamma^{eiv}=&\frac{\gamma^m}{\lambda}, \\  
    \label{eqn:beta_est}
    \beta^{eiv}=&\beta^m-\gamma^m \delta^m \Bigg(\frac{1-\lambda}{\lambda}\Bigg).
\end{align}
All the elements on the right-hand side are known under \Cref{ass:classical}, except for the reliability ratio. 
The EIV strategy then plugs a reliable estimate for $\lambda$ into \eqref{eqn:gamma_est} and \eqref{eqn:beta_est} to estimate the parameters of interest. 

\subsubsection{The first stage}

\Cref{prop:iv} shows that the first stage can be used as an estimate for $\lambda$ under \Cref{ass:fs}. The intuition for this result follows more easily in a  univariate setting without $SES$ as an independent variable. The univariate reliability ratio is defined by replacing $\sigma_{u_{t}}^2$ in $\lambda$ with $\sigma_{s_{t}}^2$. \Cref{ass:fs}(b) implies that $\mathrm{Cov}[s_t,s_{t-1}]=\sigma_{s_{t}}^2$, so the coefficient from the univariate first stage regression of $s^m_{t}$ on $s^m_{t-1}$ is proportional to the univariate reliability ratio. This result extends to the multivariate setting since together \Cref{ass:fs}(a)-(b) imply that $\mathrm{Cov}[u_t,u_{t-1}]=\sigma_{u_{t}}^2$.

Our EIV first stage (EIV FS) strategy uses the first stage as an estimate for $\lambda$. The following proposition formalizes what EIV FS identifies under \Cref{ass:classical} and \ref{ass:fs}.
\begin{proposition}\label{prop:fs}
Under \Cref{ass:classical} and \ref{ass:fs}, and \Cref{eqn:shrinking}, it holds that: 
    \begin{align}
        \gamma^{eiv}=&\gamma \Bigg(\frac{1}{\widetilde{\lambda}}\Bigg), \\ 
        \beta^{eiv}=&\beta.
    \end{align}
Where $\gamma^{eiv}$ and $\beta^{eiv}$ are defined by replacing $\lambda$ with $\pi'$ in \eqref{eqn:gamma_est} and \eqref{eqn:beta_est}.    
\end{proposition}

Similar to the IV strategy, EIV FS corrects attenuation bias in the test score coefficient under classical measurement error, overestimates it under non-classical error, and addresses contamination bias in the SES coefficient in both cases. Unlike the IV strategy, EIV FS achieves these results under the weaker \Cref{ass:fs}. It relies directly on the medium regression estimate, whereas the IV strategy uses the reduced form estimate. The two are equivalent only if $\mathrm{Cov}[e_{t},\Delta_{s_t}]=0$, which 
holds when $\Delta_{s_t}$ is constant. Whereas \Cref{ass:rf} only seems likely when the two tests are administered arbitrarily shortly after one another, \Cref{ass:fs} appears generally more plausible. In particular, under \Cref{ass:fs} the teachers may pay attention to $\Delta_{s_t}$ in their subjective evaluations, so that it is part of the error term $e_{t}$, but $\Delta_{s_t}$ cannot correlate with the included variable $SES$ and unobserved ability $s_t$. 

\subsubsection{Test score history}

We propose a novel EIV test score history (EIV TH) strategy that recovers the reliability ratio under \Cref{ass:classical} only, while making use of data on students' test score histories. The intuition of this strategy is to approximate the ideal conditions of the test-retest method, which estimates the reliability ratio using a test administered at time $t-\epsilon$, arbitrarily short before time $t$ \citep{Cito2017}. With this ideal test from period $t-\epsilon$, the first stage equation is,
\begin{align}
    s_{t}^m=\kappa_{t-\epsilon}SES +\pi_{t-\epsilon} s_{t-\epsilon}^m+e_t^{fs_{\epsilon}},
\end{align}
where we include a time subscript on the coefficients. For a test administered arbitrarily close to time $t$, \Cref{ass:rf} holds trivially, and \Cref{prop:iv} shows that $\pi_{t-\epsilon}'= \lambda$. While a test score that close to time $t$  may not be available, we may observe multiple test scores further away. It is plausible that such scores satisfy \Cref{ass:classical}, even if \Cref{ass:fs} and \ref{ass:rf} may not hold. Nevertheless, the EIV TH strategy leverages these test scores observed at earlier points in time to identify the reliability ratio.

We begin by repeating the first stage equation, sequentially replacing the test score from period $t-1$ on the right-hand side with scores from earlier periods ,
\begin{align}\label{eqn:fs_tau}
    s_{t}^m= \kappa_{t-\tau}SES + \pi_{t-\tau} s_{t-\tau}^m+e_t^{fs_{\tau}}, \quad \forall \tau \geq 1.
\end{align}
It follows from  the proof of \Cref{prop:iv} that under \Cref{ass:classical} only, $\pi_{t-\tau}'$ is a biased estimate for the reliability ratio,
\begin{align}
   \pi_{t-\tau}'=& \lambda B_{\pi_{t-\tau}}, \quad \forall \tau \geq 1,
\end{align}
with $B_{\pi_{t-\tau}}=\Bigg( \frac{\mathrm{Cov}[u_{t},u_{t-\tau}]}{\sigma_{u_{t}}^2}\Bigg)$. Therefore, we use $\pi_{t-\tau}'$ as a dependent variable in a forecasting regression equation of the form,
\begin{align}\label{eqn:forecasting}
    \pi_{t-\tau}'=& f(t-\tau)+\varepsilon_{t-\tau}. 
\end{align}
Subsequently, we make a one-period out-of-sample forecast towards $\tau=\epsilon$, denoted by $ \widehat{f}(t-\epsilon)$. Without forecasting error this resembles the ideal conditions of the test-retest method, such that $\widehat{f}(t-\epsilon)={f}(t-\epsilon)= \pi_{t-\epsilon}'=\lambda$. 
In the absence of forecasting error, the chosen polynomial in \eqref{eqn:forecasting} accurately captures the changes in $\pi'_{t-\tau}$ over time. While the EIV FS strategy estimates $\gamma^{eiv}$ and $\beta^{eiv}$ by replacing $\lambda$ with $\pi'$ in \eqref{eqn:gamma_est} and \eqref{eqn:beta_est}, EIV TH estimates them by replacing $\lambda$ with ${f}(t-\epsilon)$.

\section{Results}\label{sec:Results}

The baseline OLS estimates are presented in columns (1) and (2) of \Cref{tbl:OLS_IV}. Column (1) reports estimates from a regression where the dependent variable is a dummy equal to one if the teacher’s track recommendation is at or above the college track, and the independent variables are parental income (SES) and the fifth-grade end term test score. Consistent with previous evidence from the Netherlands, the findings show that students with higher income parents are more likely to receive at least a college track recommendation, conditional on the fifth-grade end term test. The estimate for the conditional SES gap implies that a one standard deviation increase in parental income is associated with a 2.8 percentage point increase in receiving at least a college track recommendation.

Column (1) further shows that test scores are positively associated with teacher track recommendations. As our test scores are subject to classical measurement error, this estimate is attenuated. Whether and how this contaminates the OLS estimate for the conditional SES gap depends upon the relationship between test scores and SES. This relationship is presented in column (2). As expected, SES is positively associated with the fifth-grade end term test score. We conclude that OLS overestimates the conditional SES gap, where the magnitude of this positive contamination bias also depends on the reliability ratio of the test.

\Cref{tbl:OLS_controls} in Appendix D shows OLS estimates of the conditional SES gap while including additional control variables: gender, migration dummies, age, and cohort fixed effects. The estimated gap only changes marginally, from 0.028 to 0.029. It may seem unsurprising that the estimate changes little, given that most included controls correlate weakly with parental income, with migration background as a clear exception.
In general, these results suggest that our main estimates are robust to controlling for additional variables typically observed in administrative data.

\begin{table}[t!]
\begin{center}
\caption{OLS and IV estimates}
\label{tbl:OLS_IV}
\begin{tabular}{l*{5}{c}}
\toprule\toprule
&\multicolumn{2}{c}{OLS strategy}&\multicolumn{3}{c}{IV strategy}\\
&\multicolumn{1}{c}{(1)}&\multicolumn{1}{c}{(2)}&\multicolumn{1}{c}{(3)}&\multicolumn{1}{c}{(4)}&\multicolumn{1}{c}{(5)}\\
&\multicolumn{1}{c}{$\mathbbm{1}$(recom. $\geq$}&\multicolumn{1}{c}{End term}&\multicolumn{1}{c}{End term}&\multicolumn{1}{c}{$\mathbbm{1}$(recom. $\geq$}&\multicolumn{1}{c}{$\mathbbm{1}$(recom. $\geq$}\\
[-0.75em]
&\multicolumn{1}{c}{college tr.)}&\multicolumn{1}{c}{grade 5}&\multicolumn{1}{c}{grade 5}&\multicolumn{1}{c}{college tr.)}&\multicolumn{1}{c}{college tr.)}\\
\midrule 
SES income          &       0.028***&       0.229***  & 0.031***
                    &       0.030***&       0.016***\\
                    &     (0.001)   &     (0.004)   &     (0.001) &     (0.001)   &     (0.001)   \\
End term grade 5    &       0.381***&              &                                  &               &       0.432***\\
                    &     (0.001)   &                &            &               &     (0.002)   \\
Midterm grade 5     &       &        &       0.887***&       0.383***&               \\
                    &        &     &     (0.002)   &     (0.001)  &               \\              
[.25em]
Observations        &      148019   &      148019   & 148019  & 148019  & 148019  \\
\(R^{2}\)           &       0.498   &       0.064   &  0.794 & 0.500 &   0.490   \\
$\Bigg(\frac{\sigma_{u_{t-1}}^2+\sigma_{m_{t-1}}^2}{ \sigma_{u_{t}}^2+\sigma_{m_{t}}^2}\Bigg)$ &   &   &  0.992 &   &   \\
\bottomrule \bottomrule
    \end{tabular}
        \begin{tablenotes}
	\footnotesize
        \item[] Notes:  ***, **, * refers to statistical significance at the 1, 5, and 10\% level. Coefficients in columns (1) through (4) are estimated with OLS, and in column (5) with 2SLS. Standard errors (in parentheses) are clustered at the school level.
        \end{tablenotes}
 \end{center}
\end{table}

\subsection{The IV strategy}

Columns (3) through (5) of \Cref{tbl:OLS_IV} show the results of the IV strategy, which uses the fifth-grade midterm test as an instrument for the fifth-grade end term test. The first stage in column (3) shows that the estimated coefficient on the midterm test is equal to $0.887$. This estimate measures the reliability ratio after multiplication with the fraction containing the variances of the error terms from the balancing regression. Hence, the first stage estimate for the reliability ratio is $0.887 \times 0.992 = 0.880$.

The IV strategy addresses measurement error by using the first stage to correct the reduced form estimate on the midterm term test shown in column (4). The second stage estimate on the end term test in column (5) can be calculated by dividing the reduced form by the first stage, $\frac{0.383}{0.887}=0.432$. The corresponding IV estimate for the conditional SES gap is equal to 0.016, which implies that a one standard deviation increase in parental income is associated with a 1.6 percentage point increase in receiving at least a college track recommendation.

\subsection{The EIV strategy}

The EIV strategy requires a value for the reliability ratio of the end term test. From the IV strategy we know that the first stage estimate for the reliability ratio is 0.880. Reliability ratios near 0.90 are considered high \citep{Cito2017}. Depending on the topic, the GMAT is advertised to have a reliability ratio between 0.89 and 0.90 \citep{GMAT2023}, the GRE between 0.87 and 0.95 \citep{ETS2023}, and the SAT between 0.89 and 0.93 \citep{collegeboard2013}. 

The EIV FS strategy uses the first stage as an estimate for the reliability ratio to directly address the bias in the OLS estimates. In particular, with the reliability ratio equal to $0.880$, the OLS estimate on the fifth-grade end term test is attenuated by a factor of $0.880$, and the conditional SES gap is overestimated by a value of $\gamma^m \delta^m \big(\frac{1-\lambda}{\lambda}\big)=0.381\times0.229\times\big(\frac{1-0.880}{0.880}\big)=0.012$. Hence, the EIV FS estimates are equal to $\big(\frac{0.381}{0.880}\big)=0.433$ and $0.028-0.012=0.016$, respectively. 

\begin{figure}[t!]
\centering
\caption{Estimate for the reliability ratio using students' test score histories}
\label{fig:TS_pitilde} 
\includegraphics[width=.70\textwidth]{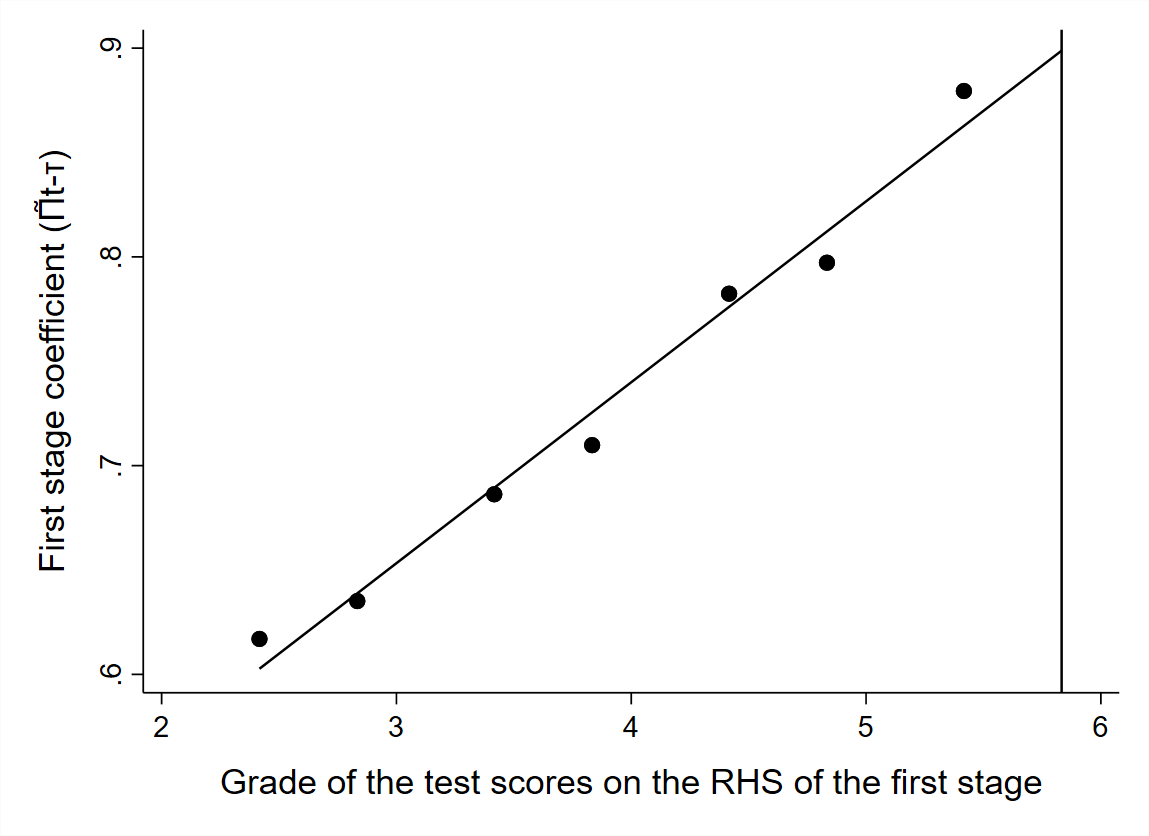} 
\caption*{ \footnotesize
Notes: this figure plots the estimates for ${\pi}'_{t-\tau}$ and the forecasting regression used to estimate $\lambda$. 
We use OLS to estimate $\pi_{t-\tau}$ from \eqref{eqn:fs_tau} and the variance of the error term from \eqref{eqn:balancing}, to obtain ${\pi}'_{t-\tau}$ for each $\tau$. The forecasting regression equation \eqref{eqn:forecasting} is specified as a polynomial of degree one and estimated with OLS.
}
\end{figure}

The novel EIV TH strategy uses an alternative estimate for the reliability ratio, which is obtained by using the biannual test score data throughout grade two to five. \Cref{fig:TS_pitilde} presents the estimates from the first stage regressions that separately use the midterm and end term tests from grade two to five as independent variable. Instead of using the rightmost estimate on the fifth-grade midterm test as the reliability ratio, the EIV TH approach relies on the first stage estimates of all previous test scores to predict the first stage estimate of a test score  very close to the fifth-grade end term test. This approximates the estimate obtained under the ideal conditions of the test-retest method. We use a linear polynomial for the forecasting regression equation. With an $R^2$ of $0.981$ this provides a good fit for the development of the first stage estimates. The estimate for the reliability ratio is the one-period out-of-sample forecast and is equal to $0.899$. The EIV TH strategy produces an estimate for the fifth-grade end term test of 0.424 and an estimate for the conditional SES gap of $0.018$. 

\subsection{Comparing results}

\begin{figure}[t!]
\centering
\caption{OLS, IV, and EIV estimates}
\label{fig:OLSIVEIV} 
   \includegraphics[width=.90\textwidth]{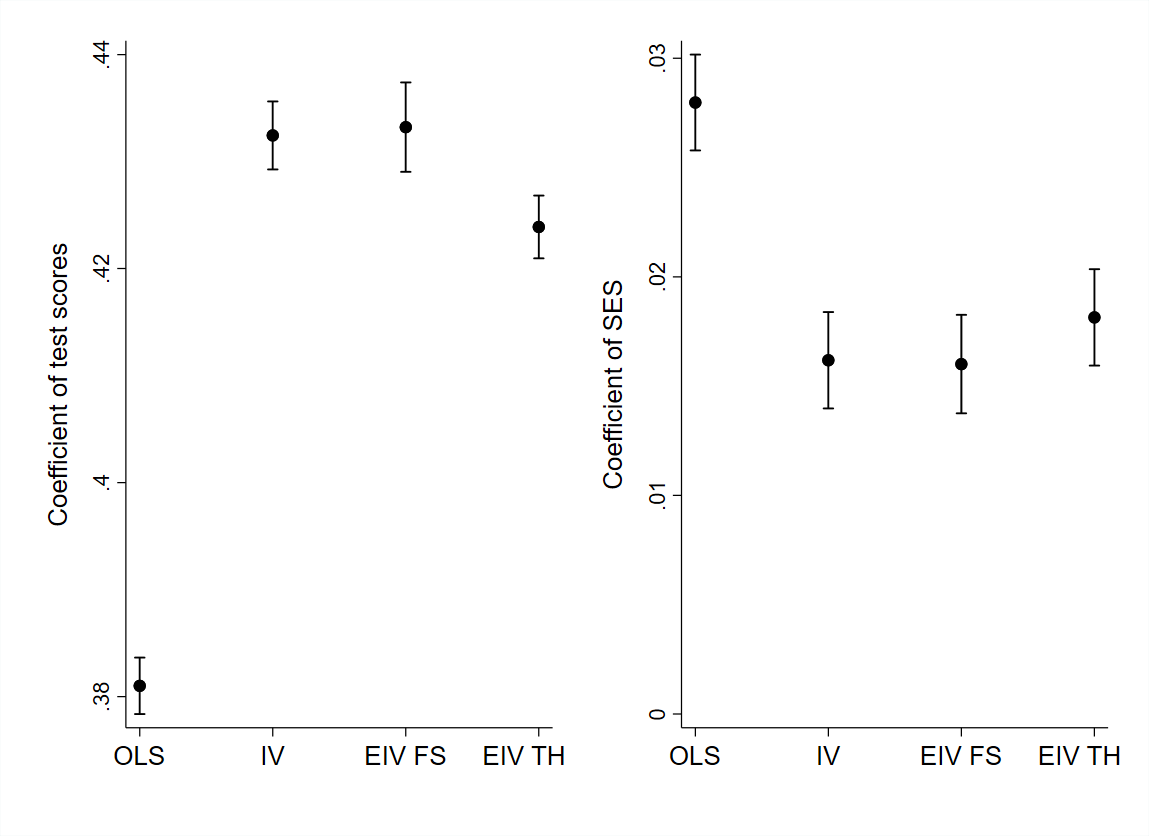}
\caption*{ \footnotesize
Notes: this figure shows the estimated coefficients for the OLS, IV, and EIV strategy. The lines are 95\% confidence intervals. The estimates and standard errors for the OLS and IV strategy can be found in \Cref{tbl:OLS_IV}.
The estimates and standard errors for EIV FS are obtained via a stacking approach. We triplicate the data and estimate \eqref{eqn:med}, \eqref{eqn:balancing}, and \eqref{eqn:fs} with OLS via a single regression. We cluster standard errors at the school id that is similarly triplicated.
For EIV TH we use the same stacking approach to estimate \eqref{eqn:med} and \eqref{eqn:balancing}, while $\lambda$ is estimated from the forecasting regression equation \eqref{eqn:forecasting} (see \Cref{fig:TS_pitilde} for details) and considered fixed. 
}
\end{figure}

\Cref{fig:OLSIVEIV} shows the estimates with 95\% confidence intervals for the OLS strategy together with all three strategies that aim to address measurement error. The left figure presents the estimates for the end term test and visualizes the attenuation bias of OLS with classical measurement error. The IV and EIV FS strategy inflate the test score estimate by a comparable amount, since the reduced form estimate on the midterm test scores ($0.383\times 0.992=0.380$) is similar to the medium regression estimate on the end term test scores ($0.381$). The right figure presents the estimates for the conditional SES gap. Similar attenuation bias translates into a similar contamination bias, and so the IV and EIV FS strategy also produce essentially the same estimate for the conditional SES gap. The attenuation and contamination bias are somewhat smaller for the EIV TH strategy, since the estimate for the reliability ratio is somewhat higher when using the one-period out-of-sample forecast.

The formula  $\Bigg(\frac{\beta^m-\beta^{\bullet}}{\beta^m} \Bigg) \times 100$ quantifies the share of the conditional SES gap in track recommendations attributable to measurement error. Here, $\beta^m$ is the OLS estimate and $\beta^{\bullet}$ refers to the IV and EIV estimates. The IV strategy attributes 42.14\% of the SES gap to measurement error, while the EIV strategies yield estimates ranging from 35.13\% to 42.78\%, depending on how the reliability ratio is estimated. Overall, the results highlight the importance of addressing contamination bias and that, in our setting, the three strategies produce broadly consistent findings.

\subsection{Testing assumptions}

\Cref{ass:fs}(a) and \ref{ass:rf} have direct testable implications. Define $\Delta_{s^m_{t}}=s_{t}^m-s_{t-1}^m$ as the change in grade 5 test scores, and let $\alpha_x = \big(\frac{\mathrm{Cov}[\Delta {s^m_{t}},x]}{\sigma_{x}^2}\big)$ be the OLS estimator of a model that regresses $\Delta_{s^m_{t}}$ upon a single student characteristic $x$. The classical measurement error in our test scores, when they appear on the left-hand side, does not affect the OLS estimate. Hence, \Cref{ass:fs}(a) requires that $\alpha_{SES}$ is zero and \Cref{ass:rf} requires that $\alpha_{x}$ is zero for every student characteristic $x$, including $SES$.

\Cref{tbl:OLS_ass2and3} in Appendix D shows OLS estimates of the change in grade 5 test scores on SES and all additional control variables. Column (1) shows that the estimate of SES is statistically significant at the 1\% level. Although we can reject \Cref{ass:fs}(a), the SES estimate of 0.005 seems economically small. In particular, the estimate is reduced by 97.82\% compared to the SES estimate from the regression of the fifth-grade end term test score on SES in \Cref{tbl:OLS_IV}. Column (2) to (5) further show that the gender dummy, the immigrant dummies, and age also correlate with the change in test scores at the 1\% significance level. The coefficients for the immigrant dummies seem of non-negligible size. These results cast doubt on \Cref{ass:rf}. In general, \Cref{ass:rf} seems unlikely in settings where the difference between time period $t$ and $t-1$ is large, especially if the variable is highly dynamic.

Though it is difficult to empirically test \Cref{ass:fs}(b), one may discuss its implications for the change in ability over time. In particular, if $\Delta_{s_t}$ is uncorrelated with $s_t$, the following two equalities directly follow:
\begin{align}
\label{eqn:overtime_2}
-\sigma_{\Delta{s_{t}}}^2=& \mathrm{Cov}[\Delta_{s_t},s_{t-1}], \\ 
\label{eqn:overtime_3}
 \sigma_{\Delta{s_{t}}}^2= & \sigma_{s_{t-1}}^2-\sigma_{s_{t}}^2.
\end{align}
Low ability students in $t-1$ must have had larger positive changes in ability than high ability students in $t-1$ (from \eqref{eqn:overtime_2}) and the variance of ability must be larger in $t-1$ than in $t$ (from \eqref{eqn:overtime_3}). Whether these equalities are likely depends on the setting at hand. In an education context, \eqref{eqn:overtime_2} and \eqref{eqn:overtime_3}  are consistent with diminishing marginal returns to ability.

The EIV TH strategy also relies, in addition to \Cref{ass:classical}, on an accurate prediction of changes in the first stage estimate over time. \Cref{fig:TS_pitilde} supports this, showing that a linear polynomial explains 98.1\% of the variation in the first stage estimates. This also suggests that the factors driving the first stage away from the reliability ratio, which are $\mathrm{Cov}[\Delta{s_{t}},SES]\neq 0$ and $\mathrm{Cov}[\Delta{s_{t}},s_{t}]\neq 0$, diminish over time in a predictable way. This is further supported by results presented in \Cref{fig:Assumption_TS} in Appendix D, which plots OLS estimates of $\Delta{s^m_{t}}$ on $SES$ across time and shows that $\mathrm{Cov}[\Delta{s^m_{t}},SES]$ converges to zero in an approximately linear fashion.

\subsection{Robustness checks}

We test the robustness of our findings in four ways. First, \Cref{fig:OLS_IV_EIV_recom} in Appendix D shows the results when using two alternative measures for the teacher track recommendation: a dummy that equals one if the teacher recommendation is equal to or higher than the theory-oriented vocational track (\Cref{fig:OLS_IV_EIV_recom}(a)) or the university track (\Cref{fig:OLS_IV_EIV_recom}(b)). Second, \Cref{fig:OLS_IV_EIV_ses} in Appendix D shows the results when using three alternative SES measures: a dummy that equals one if parental income is above the median of a student's cohort (\Cref{fig:OLS_IV_EIV_ses}(a)), years of schooling of the highest educated parent (\Cref{fig:OLS_IV_EIV_ses}(b)), and a dummy that equals one if the highest educated parent completed at least college education (\Cref{fig:OLS_IV_EIV_ses}(c)). Third, \Cref{fig:OLS_IV_EIV_cohorts} in Appendix D shows the results when separately estimating the models per cohort (\Cref{fig:OLS_IV_EIV_cohorts}(a)-(d)). All results are similar to our baseline findings.

Fourth, as SES is strongly correlated with migration background, a potential concern is that the conditional SES gap at least partly reflects a gap by migration background. To test whether this is the case, we estimated our baseline models for immigrants and natives separately. \Cref{fig:OLS_IV_EIV_imm} in Appendix D shows that the SES estimates are similar in both subgroups.

\section{Conclusion}\label{sec:Conclusion}

Teacher bias is typically assessed by examining  gaps in teacher's subjective evaluations, conditional on objective measures of student abilities. A key challenge in identifying such bias is the presence of measurement error in test scores, which are used as proxies for student ability. We show that both classical and certain forms of non-classical measurement error introduce equivalent bias in the estimated conditional gap in subjective evaluations. We discuss three different strategies to address this contamination bias. While an IV approach is more commonly used in applied econometric studies, we demonstrate that an EIV approach can address measurement error under weaker assumptions.

The empirical analysis focuses on conditional SES gaps in teacher track recommendations in the Netherlands. This is a setting where potential teacher bias is highly consequential and findings on conditional gaps with respect to parental SES are widely accepted. Using new administrative data, we find that measurement error in test scores can explain a substantial portion of the conditional SES gap, between 35 and 43\%.

It may seem striking that measurement error explains such a large share of the conditional SES gap, especially since our test scores are reliable by conventional norms. Yet when student ability is strongly correlated with SES and matters for teacher track recommendations, contamination bias can still be substantial. While our findings show that teacher bias may be smaller than previously reported, the strong correlation between ability and SES points to significant inequalities of opportunity before and during primary education.

\addcontentsline{toc}{section}{References}
\bibliographystyle{chicagoa}
\bibliography{Library}

\newpage

\appendix

\section{\texorpdfstring{Proof \Cref{prop:ols}}{proof}}\label{sec:proof_ols}

We start with $\delta^m$ from the balancing regression equation \eqref{eqn:balancing}. Using \Cref{ass:classical} we derive 
\begin{align}
    \delta^m=&\Bigg(\frac{\mathrm{Cov}[\widetilde{\lambda} (s_t+m_t), SES]}{\mathrm{Var}[SES]}\Bigg)= \Bigg(\frac{\mathrm{Cov}[ s_t+m_t, SES]} {\mathrm{Var}[SES]}\Bigg)\widetilde{\lambda} = \Bigg(\frac{\mathrm{Cov}[ s_t, SES]} {\mathrm{Var}[SES]}\Bigg) \widetilde{\lambda} \\ \notag
    =& \delta \widetilde{\lambda}. 
\end{align}
Hence, we can write the balancing regression equation as
\begin{align}
 \begin{pmatrix} s_t \\ s_{t}^m  \end{pmatrix}= \begin{pmatrix} s_t \\ \widetilde{\lambda} (s_t+m_t)  \end{pmatrix}= 
\begin{pmatrix} \delta \\ \widetilde{\lambda} \delta \end{pmatrix} SES +  \begin{pmatrix} u_{t} \\ \widetilde{\lambda}(u_{t}+m_t) \end{pmatrix}.
\end{align} 
Applying the Frisch-Waugh logic and \Cref{ass:classical}, we can now derive the expression for $\gamma^m$ as follows,
\begin{align}
    \gamma^m=&\Bigg(\frac{\mathrm{Cov}[y_t,u^m_t]}{\mathrm{Var}[u^m_t]}\Bigg)=\Bigg(\frac{\mathrm{Cov}[y_t, \widetilde{\lambda}(u_{t}+m_t)]}{\mathrm{Var}[ \widetilde{\lambda}(u_{t}+m_t)]}\Bigg)=\Bigg(\frac{\mathrm{Cov}[y_t, (u_{t}+m_t)]}{\mathrm{Var}[ u_{t}+m_t]}\Bigg) \Bigg(\frac{\widetilde{\lambda}}{\widetilde{\lambda}^2}\Bigg) \\ \notag
    =& \Bigg(\frac{\mathrm{Cov}[y_t,u_{t}]}{\sigma_{u_{t}}^2+\sigma_{m_{t}}^2}\Bigg)\Bigg(\frac{1}{\widetilde{\lambda}}\Bigg)= \Bigg(\frac{\mathrm{Cov}[y_t,u_{t}]}{\sigma_{u_{t}}^2} \Bigg) \Bigg(\frac{\sigma_{u_{t}}^2}{\sigma_{u_{t}}^2+\sigma_{m_{t}}^2} \Bigg) \Bigg(\frac{1}{\widetilde{\lambda}}\Bigg) \\ \notag
    =&\gamma \Bigg(\frac{\lambda}{\widetilde{\lambda}}\Bigg), 
\end{align}
where $\lambda=\Bigg(\frac{\sigma_{u_{t}}^2}{\sigma_{u_{t}}^2+\sigma_{m_{t}}^2} \Bigg)$.
To derive the expression for $\beta^m$, we first introduce the short regression equation:
\begin{align}\label{eqn:short}
    y_t=\beta^s SES+e_t^s.
\end{align}
Next, we use the omitted variable bias formula for $\beta^s$ to write
\begin{align}
\beta^{m}+\gamma^{m}\delta^{m}=\beta+\gamma\delta.
\end{align}
Rewriting for $\beta^m$, substituting for $\delta^{m}=\delta\widetilde{\lambda}$ and  $\gamma^{m}=\gamma \Bigg(\frac{\lambda}{\widetilde{\lambda}
}\Bigg)$, and simplifying yields
\begin{align}
\beta^{m}=&\beta+\gamma\delta\bigg(1-\lambda\bigg).
\end{align}

\section{\texorpdfstring{Proof \Cref{prop:iv}}{proof}}\label{sec:proof_iv}

We first derive results using only \Cref{ass:classical}. Subsequently, we simplify these results by additionally invoking \Cref{ass:fs} and \ref{ass:rf} in Appendix Sections B.1 and B.2, respectively.

We start with $\delta^m_{t-1}$ from balancing regression equation \eqref{eqn:balancing_tmin1}. Using \Cref{ass:classical} we derive
\begin{align}
    \delta^m_{t-1}=&\Bigg(\frac{\mathrm{Cov}[\widetilde{\lambda} (s_{t-1}+m_{t-1}), SES]}{\mathrm{Var}[SES]}\Bigg) \\ \notag
    =& \delta_{t-1} \widetilde{\lambda}. 
\end{align}
Hence, we can write the balancing regression equation as
\begin{align}
 \begin{pmatrix} s_{t-1} \\ s_{t-1}^m \end{pmatrix}= \begin{pmatrix} s_{t-1} \\ \widetilde{\lambda}(s_{t-1}+m_{t-1}) \end{pmatrix}= \begin{pmatrix} \delta_{t-1} \\ \widetilde{\lambda}\delta_{t-1}  \end{pmatrix} SES +  \begin{pmatrix}  u_{t-1}  \\  \widetilde{\lambda} (u_{t-1} + m_{t-1}) \end{pmatrix}.
\end{align} 
Applying the Frisch-Waugh logic and \Cref{ass:classical}, we can now derive the expression for the first stage coefficient $\pi$ as follows,
\begin{align}\label{eqn:fs_noass}
    \notag
    \pi=&\Bigg(\frac{\mathrm{Cov}[s_{t}^m,u^m_{t-1}]}{\mathrm{Var}[u^m_{t-1}]}\Bigg)=
    \Bigg(\frac{\mathrm{Cov}[\widetilde{\lambda}(s_{t}+m_{t}),\widetilde{\lambda} (u_{t-1}+m_{t-1})]}{\mathrm{Var}[\widetilde{\lambda} (u_{t-1}+m_{t-1})]} \Bigg)  =\Bigg(\frac{\mathrm{Cov}[s_{t},u_{t-1}]}{\sigma_{u_{t-1}}^2+\sigma_{m_{t-1}}^2} \Bigg) \Bigg(\frac{\widetilde{\lambda}\widetilde{\lambda}}{\widetilde{\lambda}^2}\Bigg)\\
    =& \Bigg(\frac{\mathrm{Cov}[s_{t},u_{t-1}]}{\sigma_{u_{t}}^2+\sigma_{m_{t}}^2}\Bigg) \Bigg( \frac{\sigma_{u_{t}}^2+\sigma_{m_{t}}^2}{\sigma_{u_{t-1}}^2+\sigma_{m_{t-1}}^2}\Bigg) \\ \notag
     =& \Bigg(\frac{\sigma_{u_{t}}^2}{\sigma_{u_{t}}^2+\sigma_{m_{t}}^2}\Bigg)\Bigg( \frac{\mathrm{Cov}[s_{t},u_{t-1}]}{\sigma_{u_{t}}^2}\Bigg) \Bigg( \frac{\sigma_{u_{t}}^2+\sigma_{m_{t}}^2}{\sigma_{u_{t-1}}^2+\sigma_{m_{t-1}}^2}\Bigg) \\ \notag
    =& \Bigg(\frac{\sigma_{u_{t}}^2}{\sigma_{u_{t}}^2+\sigma_{m_{t}}^2}\Bigg)\Bigg( \frac{\mathrm{Cov}[\delta SES + u_{t},u_{t-1}]}{\sigma_{u_{t}}^2}\Bigg) \Bigg( \frac{\sigma_{u_{t}}^2+\sigma_{m_{t}}^2}{\sigma_{u_{t-1}}^2+\sigma_{m_{t-1}}^2}\Bigg)  \\ \notag
    =& \lambda \Bigg( \frac{\mathrm{Cov}[u_{t},u_{t-1}]}{\sigma_{u_{t}}^2}\Bigg) \Bigg( \frac{\sigma_{u_{t}}^2+\sigma_{m_{t}}^2}{\sigma_{u_{t-1}}^2+\sigma_{m_{t-1}}^2}\Bigg).
\end{align}
Next rewrite the second stage equation \eqref{eqn:ss} to obtain the reduced form regression equation,  
\begin{align}\label{eqn:rf}
    y_t= \theta_0 SES +  \theta_1 s_{t-1}^m+e_t^{iv},
\end{align}
where $\frac{\theta_1}{\pi}=\frac{\gamma^{iv} \pi}{\pi}=\gamma^{iv}$. Applying the Frisch-Waugh logic and \Cref{ass:classical}, we can similarly derive an expression for the reduced form coefficient $\theta_1$,
\begin{align}\label{eqn:rf_noass}
    \theta_1=&\Bigg(\frac{\mathrm{Cov}[y_{t},u^m_{t-1}]}{\mathrm{Var}[u^m_{t-1}]}\Bigg)=\Bigg(\frac{\mathrm{Cov}[y_{t},\widetilde{\lambda}(u_{t-1}+m_{t-1})]}{\mathrm{Var}[\widetilde{\lambda}(u_{t-1}+m_{t-1})]}\Bigg)=
    \Bigg(\frac{\mathrm{Cov}[y_{t} ,u_{t-1}]}{\sigma_{u_{t-1}}^2+\sigma_{m_{t-1}}^2}\Bigg) \Bigg(\frac{\widetilde{\lambda}}{\widetilde{\lambda}^2}\Bigg)  \\ \notag
    =& \Bigg(\frac{\mathrm{Cov}[y_{t},u_{t-1}]}{\sigma_{u_{t}}^2}\Bigg) \Bigg( \frac{\sigma_{u_{t}}^2}{\sigma_{u_{t-1}}^2+\sigma_{m_{t-1}}^2}\Bigg) \Bigg(\frac{1}{\widetilde{\lambda}}\Bigg)  \\ \notag
     =& \Bigg(\frac{\mathrm{Cov}[y_{t},u_{t}]}{\sigma_{u_{t}}^2}\Bigg) \Bigg(\frac{\mathrm{Cov}[y_{t},u_{t-1}]}{\mathrm{Cov}[y_{t},u_{t}]}\Bigg)  \Bigg( \frac{\sigma_{u_{t}}^2}{\sigma_{u_{t}}^2+\sigma_{m_{t}}^2}\Bigg) \Bigg( \frac{\sigma_{u_{t}}^2+\sigma_{m_{t}}^2}{\sigma_{u_{t-1}}^2+\sigma_{m_{t-1}}^2}\Bigg) \Bigg(\frac{1}{\widetilde{\lambda}}\Bigg)  \\ \notag
     =& \gamma  \Bigg(\frac{\lambda}{\widetilde{\lambda}}\Bigg) \Bigg(\frac{\mathrm{Cov}[y_{t},u_{t-1}]}{\mathrm{Cov}[y_{t},u_{t}]}\Bigg) \Bigg( \frac{\sigma_{u_{t}}^2+\sigma_{m_{t}}^2}{\sigma_{u_{t-1}}^2+\sigma_{m_{t-1}}^2}\Bigg).
\end{align}
We can now combine the results in \eqref{eqn:fs_noass} and \eqref{eqn:rf_noass} to derive an expression for the second stage estimate $\gamma^{iv}$,
\begin{align}\label{eqn:gammaiv_noass}
    \gamma^{iv}=&\frac{\theta_1}{\pi}=\gamma \Bigg(\frac{1}{\widetilde{\lambda}}\Bigg)\Bigg(\frac{\mathrm{Cov}[y_{t},u_{t-1}]}{\mathrm{Cov}[y_{t},u_{t}]}\Bigg)\Bigg( \frac{\sigma_{u_{t}}^2}{\mathrm{Cov}[u_{t},u_{t-1}]}\Bigg).
\end{align}
To derive an expression for the second stage estimate $\beta^{iv}$, we first introduce the balancing regression equation for $\widehat{s}_{t}^m$,
\begin{align}\label{eqn:balancing_shat}
 \begin{pmatrix} s_t^m \\ \widehat{s}_{t}^m   \end{pmatrix}= \begin{pmatrix} s_t^m \\ s_t^m - e_{t}^{fs} \end{pmatrix}=\begin{pmatrix} \delta^m \\ \delta^{iv}  \end{pmatrix} SES +  \begin{pmatrix} u^m_{t}\\ u^{iv}_t  \end{pmatrix}.
\end{align} 
We have that $\delta^{iv}=\delta^m$,
\begin{align}
    \delta^{iv} =&\Bigg(\frac{\mathrm{Cov}[\widehat{s}_{t}^m,SES]}{\mathrm{Var}[SES]}\Bigg)=\Bigg(\frac{\mathrm{Cov}[ s_t^m - e_{t}^{fs},SES]}{\sigma_{SES}^2}\Bigg) = \Bigg(\frac{\mathrm{Cov}[s_{t}^m,SES]}{\sigma_{SES}^2}\Bigg) \\ \notag
    =&\delta^m,
\end{align}
and from \Cref{prop:ols} we know that under \Cref{ass:classical}, $\delta^{iv}=\delta^m=\delta \widetilde{\lambda}$. We use the omitted variable bias formula for $\beta^s$ from \eqref{eqn:short} to write
\begin{align}
    \beta^{iv}+\gamma^{iv}\delta^{iv}=&\beta+\gamma\delta.
\end{align}
Rewriting for $\beta^{iv}$ and substituting for $\delta^{iv}=\delta\widetilde{\lambda}$ and $\gamma^{iv}$ from \eqref{eqn:gammaiv_noass},  it follows that,
\begin{align}\label{eqn:betaiv_noass}
    \beta^{iv}=& \beta+ \gamma\delta \Bigg(1-\Bigg(\frac{\mathrm{Cov}[y_{t},u_{t-1}]}{\mathrm{Cov}[y_{t},u_{t}]}\Bigg)\Bigg( \frac{\sigma_{u_{t}}^2}{\mathrm{Cov}[u_{t},u_{t-1}]}\Bigg)\Bigg).
\end{align}

\subsection{Case (i)}

Here we use the results above derived under \Cref{ass:classical} to simplify the expressions while additionally invoking \Cref{ass:fs}.
First, we use \Cref{ass:fs} to find a link between $u_{t-1}$ and $u_{t}$. We can write,
\begin{align} \label{eqn:alpha}
    \delta =&\Bigg(\frac{\mathrm{Cov}[s_{t},SES]}{\mathrm{Var}[SES]}\Bigg)
    =\Bigg(\frac{\mathrm{Cov}[s_{t-1}+\Delta_{s_{t}},SES]}{\sigma_{SES}^2}\Bigg)  
    = \Bigg(\frac{\mathrm{Cov}[s_{t-1},SES]}{\sigma_{SES}^2}\Bigg) + \Bigg(\frac{\mathrm{Cov}[\Delta_{s_{t}},SES]}{\sigma_{SES}^2}\Bigg) \\ \notag
     =& \delta_{t-1} + \alpha,
\end{align}
with $\alpha=\Bigg(\frac{\mathrm{Cov}[\Delta_{s_{t}},SES]}{\sigma_{SES}^2}\Bigg)$. \Cref{ass:fs}(a) guarantees that $\delta=\delta_{t-1}$. This allows us to write \eqref{eqn:balancing} and \eqref{eqn:balancing_tmin1} as
\begin{align} \label{eqn:balancing_both}
 \begin{pmatrix} s_{t} \\ s_{t-1} \end{pmatrix}= \begin{pmatrix} s_{t} \\ s_{t}-\Delta{s_{t}} \end{pmatrix}= \delta SES +  \begin{pmatrix}  u_{t}  \\ u_{t}-\big(\Delta_{s_{t}}-\overline{\Delta}_{s_{t}}\big) \end{pmatrix}.
\end{align} 
The mean of  $\Delta_{s_{t}}$, denoted by $\overline{\Delta}_{s_{t}}$, will be absorbed by the intercept of \eqref{eqn:balancing_tmin1}, so that $u_{t-1}=u_{t}-\big(\Delta_{s_{t}}-\overline{\Delta}_{s_{t}}\big)$. We use this to write $\mathrm{Cov}[u_{t},u_{t-1}]=\sigma_{u_{t}}^2-\mathrm{Cov}[u_{t},\Delta_{s_{t}}]$. Subsequently, \Cref{ass:fs}(a)-(b) together guarantee that $\mathrm{Cov}[u_{t},\Delta_{s_{t}}]=\mathrm{Cov}[s_t-\delta SES,\Delta_{s_{t}}]=0$. Thus \Cref{ass:fs} allows us to write $\mathrm{Cov}[u_{t},u_{t-1}]=\sigma_{u_{t}}^2$. 
\Cref{ass:fs} further allows to derive an expression for $\sigma_{u_{t-1}}^2$. In particular, we can write $\sigma_{u_{t-1}}^2=\mathrm{Var}[u_{t}-\big(\Delta_{s_{t}}-\overline{\Delta}_{s_{t}}\big)]=\mathrm{Var}[u_{t}-\Delta_{s_t}]=\sigma_{u_{t}}^2+ \sigma_{\Delta_{s_t}}^2 -2 \times \mathrm{cov}[u_{t},\Delta_{s_t}]=\sigma_{u_{t}}^2+ \sigma_{\Delta_{s_t}}^2$. In short, $\sigma_{u_{t-1}}^2=\sigma_{u_{t}}^2+ \sigma_{\Delta_{s_t}}^2$.

We can use these two results to write the first stage coefficient $\pi$ from \eqref{eqn:fs_noass} as follows,
\begin{align}\label{eqn:fs_ass2}
    \pi=&\lambda \Bigg( \frac{\sigma_{u_{t}}^2+\sigma_{m_{t}}^2}{\sigma_{u_{t}}^2+ \sigma_{\Delta_{s_t}}^2+\sigma_{m_{t-1}}^2}\Bigg).
\end{align}
For the second stage estimate $\gamma^{iv}$, we first use that $u_{t-1}=u_{t}-\big(\Delta_{s_{t}}-\overline{\Delta}_{s_{t}}\big)$ to additionally simplify the following term,
\begin{align}
    \Bigg(\frac{\mathrm{Cov}[y_{t},u_{t-1}]}{\mathrm{Cov}[y_{t},u_{t}]}\Bigg)=&  \Bigg(\frac{\mathrm{Cov}[y_{t},u_{t}]}{\mathrm{Cov}[y_{t},u_{t}]}\Bigg) - \Bigg(\frac{\mathrm{Cov}[y_{t},\Delta_{s_t}]}{\mathrm{Cov}[y_{t},u_{t}]}\Bigg) \\ \notag
     =&1-\Bigg(\frac{\mathrm{Cov}[\beta SES + \gamma s_{t}+e_{t},\Delta_{s_{t}}]}{\mathrm{Cov}[y_{t},u_{t}]}\Bigg) \\ \notag
    =&1 -\Bigg(\frac{\mathrm{Cov}[e_{t},\Delta_{s_{t}}]}{\mathrm{Cov}[y_{t},u_{t}]}\Bigg). 
\end{align}
We can now write $\gamma^{iv}$ from \eqref{eqn:gammaiv_noass} as
\begin{align}\label{eqn:gammaiv_ass2}
    \gamma^{iv}=\gamma \Bigg(\frac{1}{\widetilde{\lambda}}\Bigg)\Bigg(1 -\Bigg(\frac{\mathrm{Cov}[e_{t},\Delta_{s_{t}}]}{\mathrm{Cov}[y_{t},u_{t}]}\Bigg)\Bigg),
\end{align}
and $\beta^{iv}$ from \eqref{eqn:betaiv_noass} as,
\begin{align}\label{eqn:betaiv_ass2}
    \beta^{iv}=& \beta+\gamma\delta\Bigg(\frac{\mathrm{Cov}[e_{t},\Delta_{s_{t}}]}{\mathrm{Cov}[y_{t},u_{t}]}\Bigg).
\end{align}

\subsection{Case (ii)}

Here we use the results above derived under \Cref{ass:classical} to simplify the expressions while additionally invoking \Cref{ass:rf}.
First, we use \Cref{ass:rf} to find a link between $u_{t-1}$ and $u_{t}$. Similar to \eqref{eqn:alpha}, we write
\begin{align}
    \delta =& \delta_{t-1} + \alpha,
\end{align}
with $\alpha=\Bigg(\frac{\mathrm{Cov}[\Delta_{s_{t}},SES]}{\sigma_{SES}^2}\Bigg)$. \Cref{ass:rf} guarantees again that $\delta=\delta_{t-1}$. Moreover, we can write \eqref{eqn:balancing} and \eqref{eqn:balancing_tmin1} as
\begin{align} 
 \begin{pmatrix} s_{t} \\ s_{t-1} \end{pmatrix}= \begin{pmatrix} s_{t} \\ s_{t}-\Delta{s_{t}} \end{pmatrix}= \delta SES +  \begin{pmatrix}  u_{t}  \\ u_{t} \end{pmatrix},
\end{align} 
where the constant $\Delta{s_{t}}$ is completely absorbed by the intercept of \eqref{eqn:balancing_tmin1}, so that $u_{t-1}=u_{t}$. We use this to write $\mathrm{Cov}[u_{t},u_{t-1}]=\sigma_{u_{t}}^2$ and $\sigma_{u_{t-1}}^2=\sigma_{u_{t}}^2$.

We can use these two results to write the first stage coefficient $\pi$ from \eqref{eqn:fs_noass} as follows,
\begin{align}\label{eqn:fs_ass}
    \pi=&\lambda \Bigg( \frac{\sigma_{u_{t}}^2+\sigma_{m_{t}}^2}{\sigma_{u_{t}}^2+\sigma_{m_{t-1}}^2}\Bigg).
\end{align}
For the second stage estimate $\gamma^{iv}$, we also use that $u_{t-1}=u_{t}$ additionally implies that $\Bigg(\frac{\mathrm{Cov}[y_{t},u_{t-1}]}{\mathrm{Cov}[y_{t},u_{t}]}\Bigg)=1$. We can now write $\gamma^{iv}$ from \eqref{eqn:gammaiv_noass} as
\begin{align}\label{eqn:gammaiv_ass}
    \gamma^{iv}=& \gamma \Bigg(\frac{1}{\widetilde{\lambda}}\Bigg),
\end{align}
and $\beta^{iv}$ from \eqref{eqn:betaiv_noass} as,
\begin{align}\label{eqn:betaiv_ass}
    \beta^{iv}=& \beta.
\end{align}

\section{\texorpdfstring{Proof \Cref{prop:fs}}{proof}}\label{sec:proof_fs}

\Cref{prop:iv} shows that under \Cref{ass:classical} and \ref{ass:fs} we have that ${\pi}'=\lambda$. It immediately follows that,
\begin{align}
    \gamma^{eiv}&=\frac{\gamma^m}{\pi'} =\gamma^m \Bigg(\frac{1}{\lambda} \Bigg)=\gamma\Bigg(\frac{\lambda}{\widetilde{\lambda}}\Bigg)\Bigg(\frac{1}{\lambda} \Bigg) \\ \notag
    &=\gamma\Bigg(\frac{1}{\widetilde{\lambda}}\Bigg),
\end{align}
and that
\begin{align}
    \beta^{eiv}=&\beta^m-\gamma^m \delta^m\Bigg(\frac{1-\pi'}{\pi'}\Bigg)=\beta + \gamma\delta \big(1-\lambda \big) -\gamma^m \delta^m\Bigg(\frac{1-\lambda}{\lambda}\Bigg) \\ \notag
    =&\beta + \gamma\delta \big(1-\lambda \big) -\gamma\Bigg(\frac{\lambda}{\widetilde{\lambda}}\Bigg) \delta \widetilde{\lambda} \Bigg(\frac{1-\lambda}{\lambda}\Bigg)= \beta + \gamma\delta \big(1-\lambda \big) -\gamma\delta \big(1-\lambda \big) \\ \notag
    =& \beta.
\end{align}

\newpage

\section{Additional empirical results}\label{sec:additionalresults}

\begin{table}[h!]
\begin{center}
\caption{Descriptive statistics on the test scores across all grades}
\label{tbl:descrstats_tests}
\begin{tabular}{lccccc}
\toprule \toprule
&\multicolumn{1}{c}{(1)}&\multicolumn{1}{c}{(2)}&\multicolumn{1}{c}{(3)}&\multicolumn{1}{c}{(4)}&\multicolumn{1}{c}{(5)}\\
& Obs & Mean & SD  & Min & Max  \\
\midrule
\multicolumn{6}{c}{Panel A: Reading tests} \\
Midterm grade 2     &      114743&     139.354&      27.933&      $<$42.000&     $>$264.000\\
End term grade 2    &      126688&     146.532&      26.752&      $<$55.000&     $>$282.000\\
Midterm grade 3     &      137499&     157.415&      26.149&      $<$66.000&     $>$279.000\\
End term grade 3    &      121392&     161.807&      27.028&      $<$71.000&    $>$280.000\\
Midterm grade 4     &      142069&     173.447&      25.339&      $<$73.000&     $>$309.000\\
End term grade 4    &      111460&     182.413&      27.147&      $<$79.000&    $>$337.000\\
Midterm grade 5     &      148019&     191.618&      26.766&      $<$81.000&    $>$340.000\\
End term grade 5    &      148019&     197.279&      26.983&      $<$84.000&     $>$354.000\\
[.25em]
\multicolumn{6}{c}{Panel B: Math tests} \\
Midterm grade 2     &      132855&     167.815&      30.385&      $<$30.000&     $>$278.000\\
End term grade 2    &      132976&     189.235&      29.063&      $<$53.000&     $>$289.000\\
Midterm grade 3     &      137502&     206.446&      28.042&      $<$50.000&     $>$313.000\\
End term grade 3    &      137643&     218.366&      27.299&      $<$64.000&    $>$334.000\\
Midterm grade 4     &      141986&     230.064&      26.684&      $<$89.000&     $>$356.000\\
End term grade 4    &      123702&     241.663&      26.635&      $<$100.000&     $>$361.000\\
Midterm grade 5     &      148019&     254.142&      27.142&      $<$118.000&     $>$379.000\\
End term grade 5    &      148019&     261.959&      26.085&      $<$118.000&     $>$394.000\\
\bottomrule \bottomrule
\end{tabular}
\begin{tablenotes}
\footnotesize
\item[] Notes: this table shows the descriptive statistics for the test scores across all grades. Statistics Netherlands prevents reporting minima and maxima that are based upon less than 10 observations. Therefore, the minimum (maximum) may contain the smallest (largest) number, such that at least ten observations have values lower (higher) than that number. 
\end{tablenotes}
\end{center}
\end{table}

\begin{table}[t!]
\begin{center}
\caption{OLS estimates with control variables}
\label{tbl:OLS_controls}
\begin{tabular}{l*{3}{c}}
\toprule\toprule
&\multicolumn{1}{c}{(1)}&\multicolumn{1}{c}{(2)}&\multicolumn{1}{c}{(3)}\\
&\multicolumn{3}{c}{$\mathbbm{1}$(recom. $\geq$ college tr.)}\\
\midrule
SES income          &       0.115***&       0.028***&      0.029***\\
                    &     (0.002)   &     (0.001)   &   (0.001)   \\
End term grade 5    &               &       0.381***&       0.375***\\
                    &               &     (0.001)   &     (0.001)   \\
$\mathbbm{1}$(female)&               &               &       0.011***\\
                    &               &               &     (0.002)   \\
$\mathbbm{1}$(Western imm.)&               &               &       0.036***\\
                    &               &              &     (0.004)   \\
$\mathbbm{1}$(non-Western imm.)&               &               &       0.036***\\
                    &               &               &     (0.003)   \\
Age                 &               &               &      -0.072***\\
                    &               &              &     (0.002)   \\
$\mathbbm{1}$(Cohort$=$2019-20)        &               &               &      -0.042***\\
                    &               &               &     (0.005)   \\
$\mathbbm{1}$(Cohort$=$2020-21)       &               &              &       0.030***\\
                    &               &               &     (0.005)   \\
$\mathbbm{1}$(Cohort$=$2021-22)        &               &               &      -0.072***\\
                    &               &               &     (0.005)   \\
[.25em]
Observations        &      148019   &      148019  &      148019   \\
\(R^{2}\)           &       0.053   &       0.498   &       0.506   \\
       \bottomrule \bottomrule
    \end{tabular}
        \begin{tablenotes}
	\footnotesize
        \item[] Notes:  ***, **, * refers to statistical significance at the 1, 5, and 10\% level. Coefficients are estimated with OLS. Standard errors (in parentheses) are clustered at the school level.
        \end{tablenotes}
\end{center}
\end{table}

\begin{table}[t!]
\begin{center}
\caption{$\mathrm{Cov}[\Delta{s^m_{t}},x]$ for observable characteristics $x$}
\label{tbl:OLS_ass2and3}
\begin{tabular}{l*{5}{c}}
\toprule\toprule
&\multicolumn{1}{c}{(1)}&\multicolumn{1}{c}{(2)}&\multicolumn{1}{c}{(3)}&\multicolumn{1}{c}{(4)}&\multicolumn{1}{c}{(5)}\\
&\multicolumn{5}{c}{$\Delta_{s^m_{t}}=s^m_t-s^m_{t-1}$}\\
\midrule
SES income          &       0.005***&               &               &               &               \\
                    &     (0.001)   &               &               &               &               \\
$\mathbbm{1}$(female)&               &       0.007***&               &               &               \\
                    &               &     (0.002)   &               &               &               \\
$\mathbbm{1}$(Western imm.)&               &               &       0.018***&               &               \\
                    &               &               &     (0.004)   &               &               \\
$\mathbbm{1}$(non-Western imm.)&               &               &               &       0.012***&               \\
                    &               &               &               &     (0.004)   &               \\
Age                 &               &               &               &               &      -0.005** \\
                    &               &               &               &               &     (0.002)   \\
[.25em]
Observations        &      148019   &      148019   &      148019   &      148019   &      148019   \\
\(R^{2}\)           &       0.000   &       0.000   &       0.000   &       0.000   &       0.000   \\
       \bottomrule \bottomrule
    \end{tabular}
        \begin{tablenotes}
	\footnotesize
        \item[] Notes:  ***, **, * refers to statistical significance at the 1, 5, and 10\% level. Coefficients are estimated with OLS. Standard errors (in parentheses) are clustered at the school level.
        \end{tablenotes}
 \end{center}
\end{table}

\begin{figure}[t!]
\centering
\caption{$\mathrm{Cov}[\Delta{s^m_{t}},SES]$ over time}
\label{fig:Assumption_TS} 
\includegraphics[width=.65\textwidth]{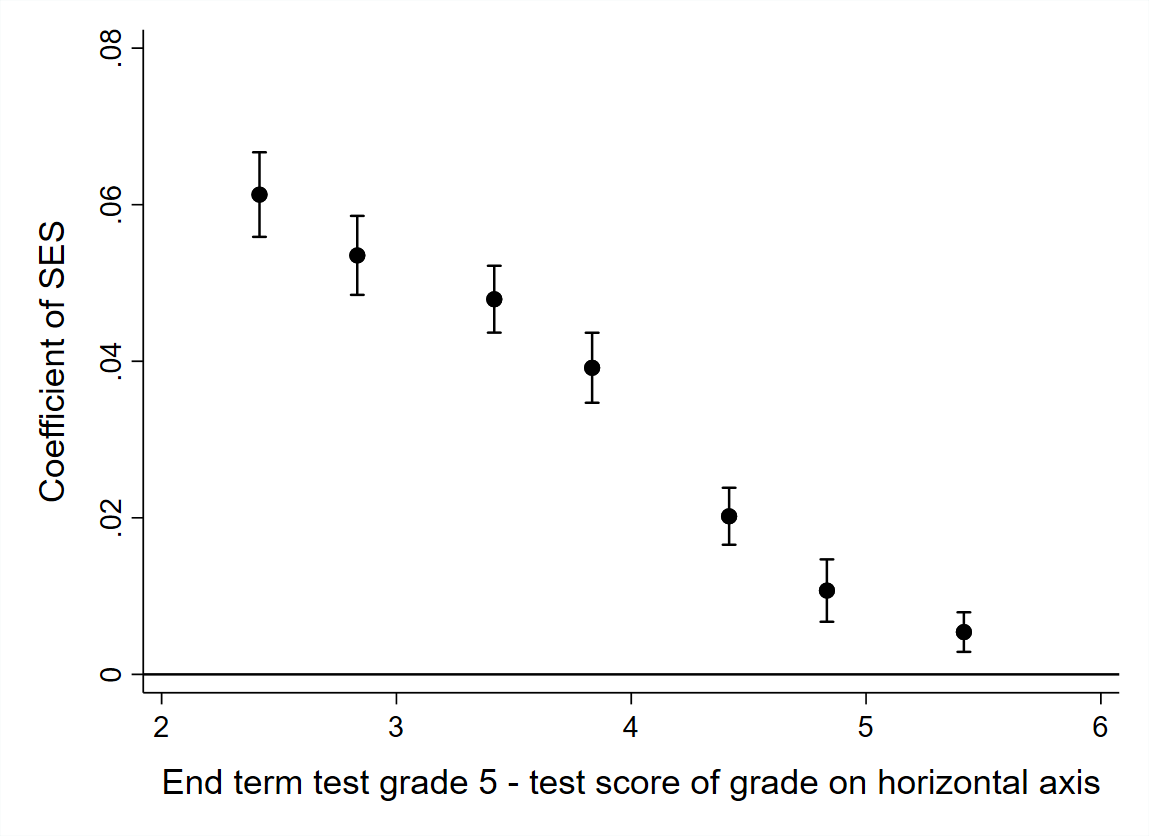} 
\caption*{ \footnotesize
Notes: this figure plots the estimates for the coefficients of SES from regressions of the change in test scores $(s^m_t-s^m_{t-\tau})$ upon SES over time $\tau$. Coefficients are estimated with OLS. Standard errors are clustered on the school id.
}
\end{figure}

\begin{figure}
\centering
\caption{OLS, IV, and EIV estimates with two alternative measures of teacher recommendation}
\label{fig:OLS_IV_EIV_recom}
\subfloat[Theory-oriented vocational track]{\label{a_rr}\includegraphics[width=.50\linewidth]{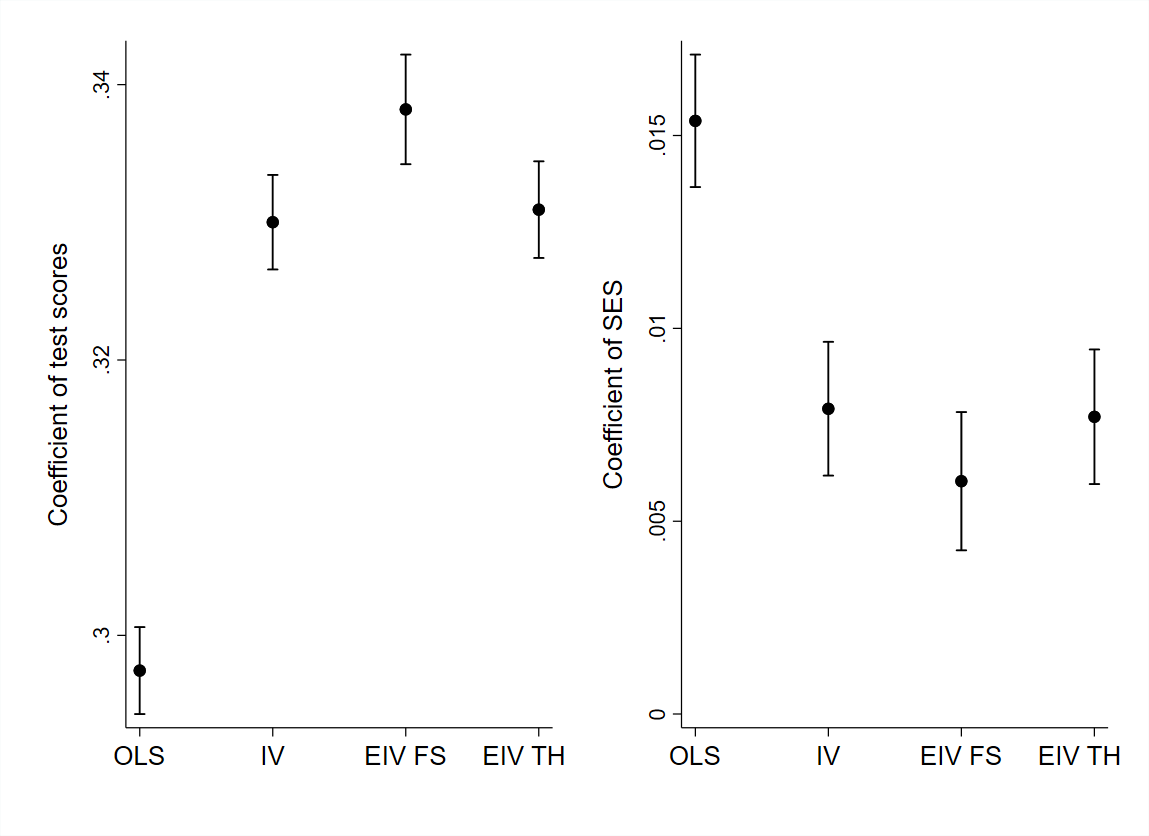}}
\subfloat[University track]{\label{b_rr}\includegraphics[width=.50\linewidth]{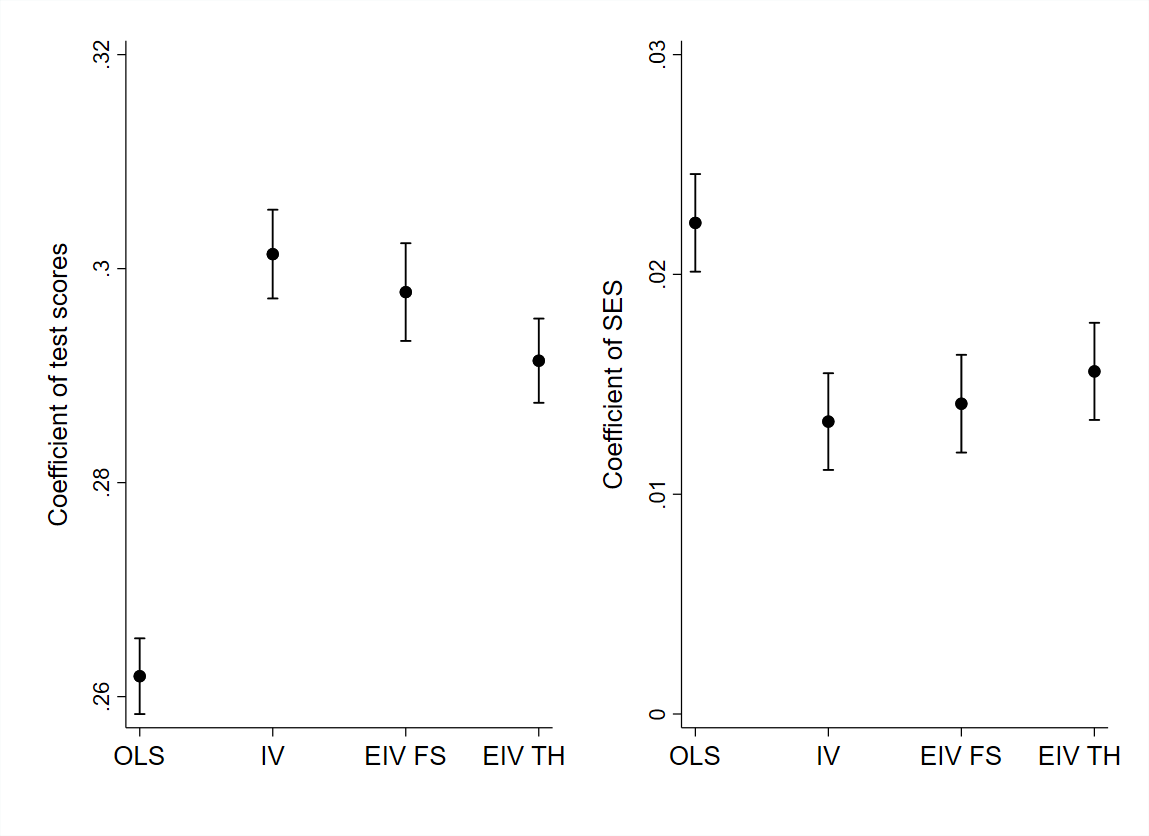}}
\caption*{ \footnotesize
Notes: this figure shows the estimated coefficients for the OLS, IV, and EIV strategy with two alternative measures of teacher recommendation. The lines are 95\% confidence intervals. The two alternative measures are a dummy variable that is equal to one if the recommendation is equal to or higher than the theory-oriented vocational track (\Cref{fig:OLS_IV_EIV_recom}(a)) or the university track (\Cref{fig:OLS_IV_EIV_recom}(b)). See \Cref{fig:OLSIVEIV} for details on the estimation.  
}
\end{figure}

\begin{figure}
\centering
\caption{OLS, IV, and EIV estimates with three alternative SES measures}
\label{fig:OLS_IV_EIV_ses}
\subfloat[Binarized income]{\label{a_rs}\includegraphics[width=.50\linewidth]{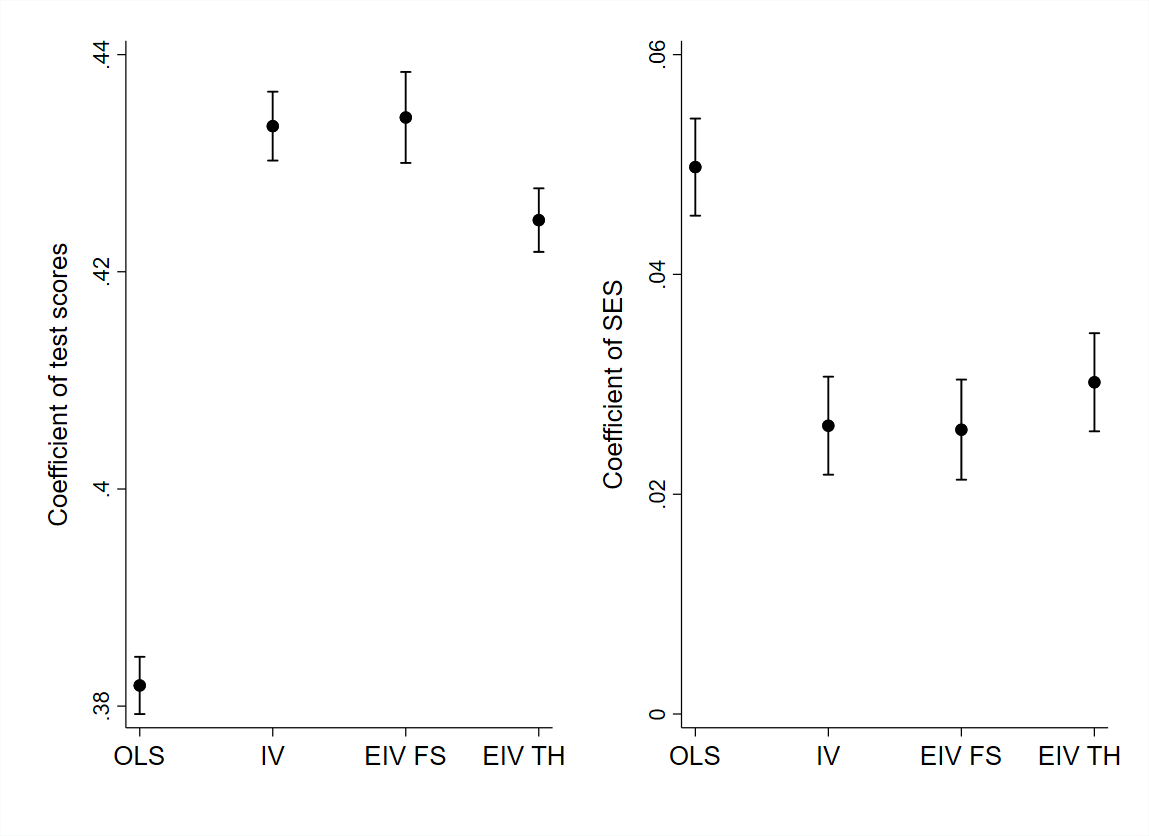}}
\subfloat[Years of schooling]{\label{b_rs}\includegraphics[width=.50\linewidth]{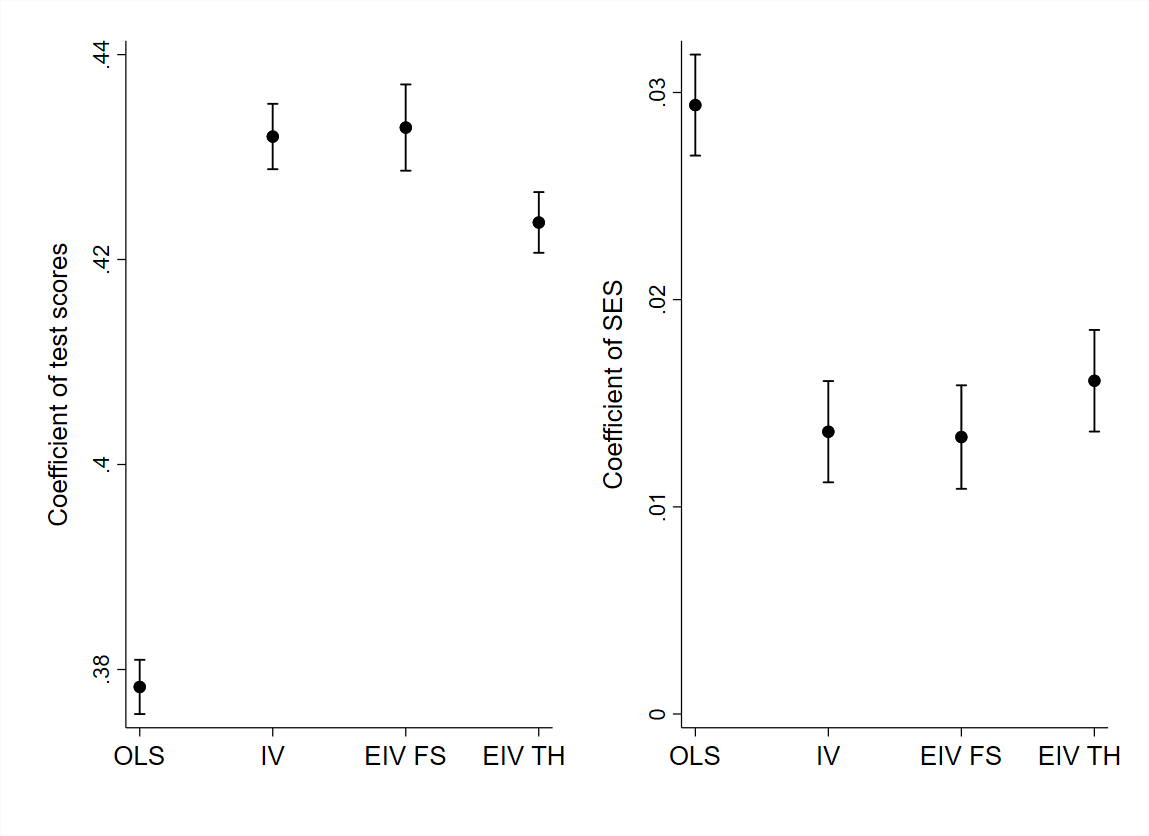}} \par
\subfloat[At least college education]{\label{c_rs}\includegraphics[width=.50\linewidth]{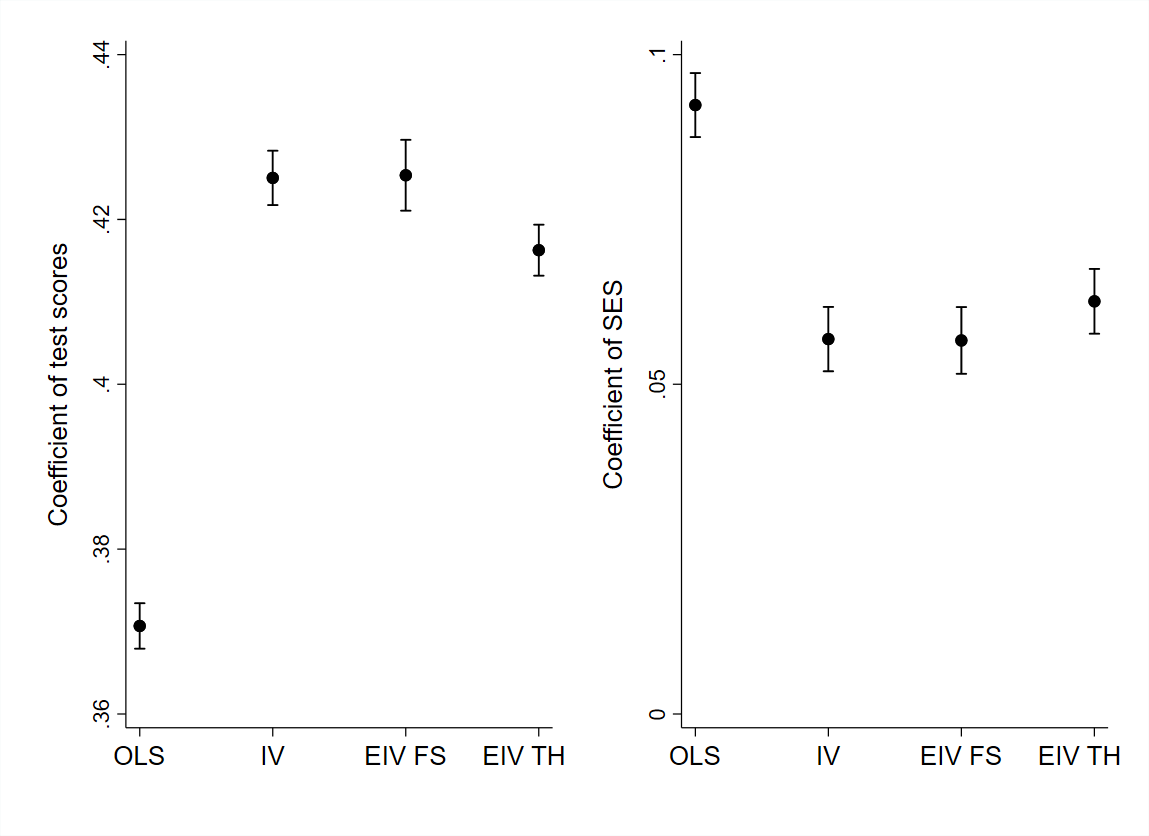}} 
\caption*{ \footnotesize
Notes: this figure shows the estimated coefficients for the OLS, IV, and EIV strategy with three alternative SES measures. The lines are 95\% confidence intervals. The three alternative measures are a dummy that equals one if parental income is above the median (\Cref{fig:OLS_IV_EIV_ses}(a)), years of schooling of the highest educated parent (\Cref{fig:OLS_IV_EIV_ses}(b)), and a dummy that equals one if the highest educated parent completed at least college education (\Cref{fig:OLS_IV_EIV_ses}(c)). See \Cref{fig:OLSIVEIV} for details on the estimation.  
}
\end{figure}

\begin{figure}
\centering
\caption{OLS, IV, and EIV estimates per cohort}
\label{fig:OLS_IV_EIV_cohorts}
\subfloat[Cohort 2018-19]{\label{a_hc}\includegraphics[width=.50\linewidth]{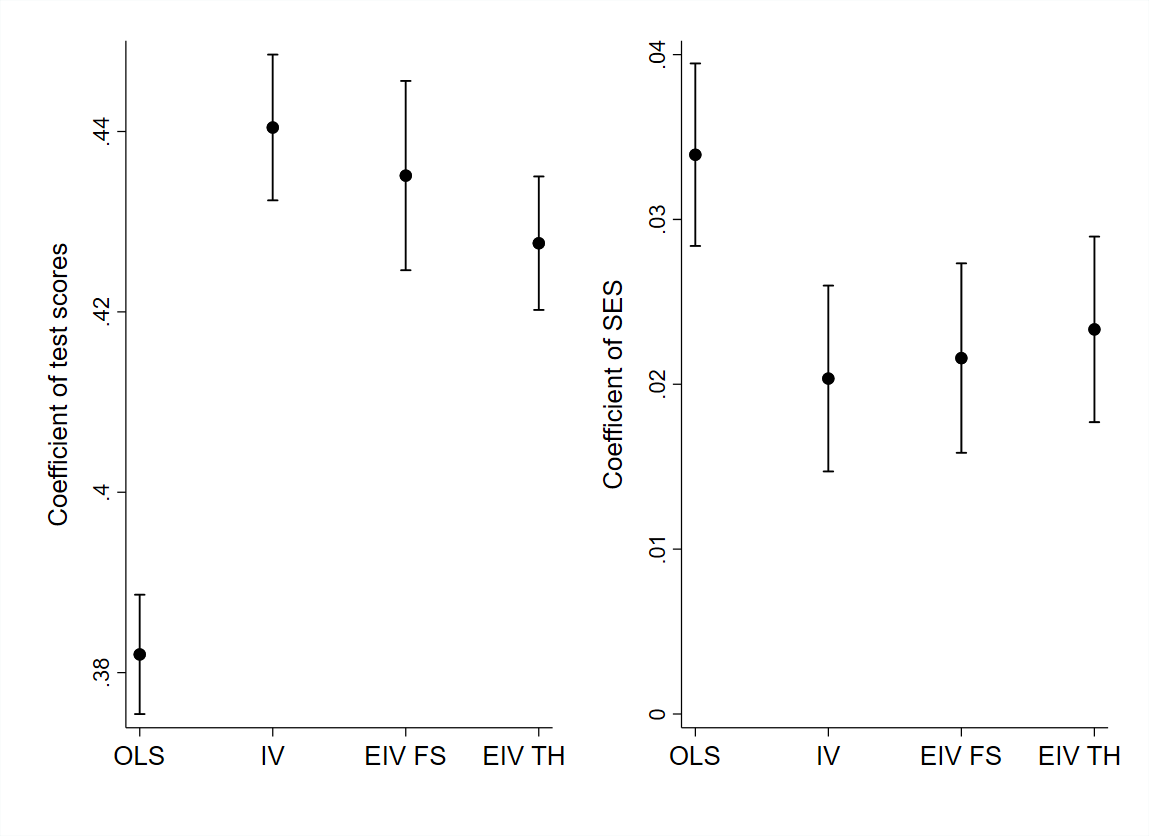}}
\subfloat[Cohort 2019-20]{\label{b_hc}\includegraphics[width=.50\linewidth]{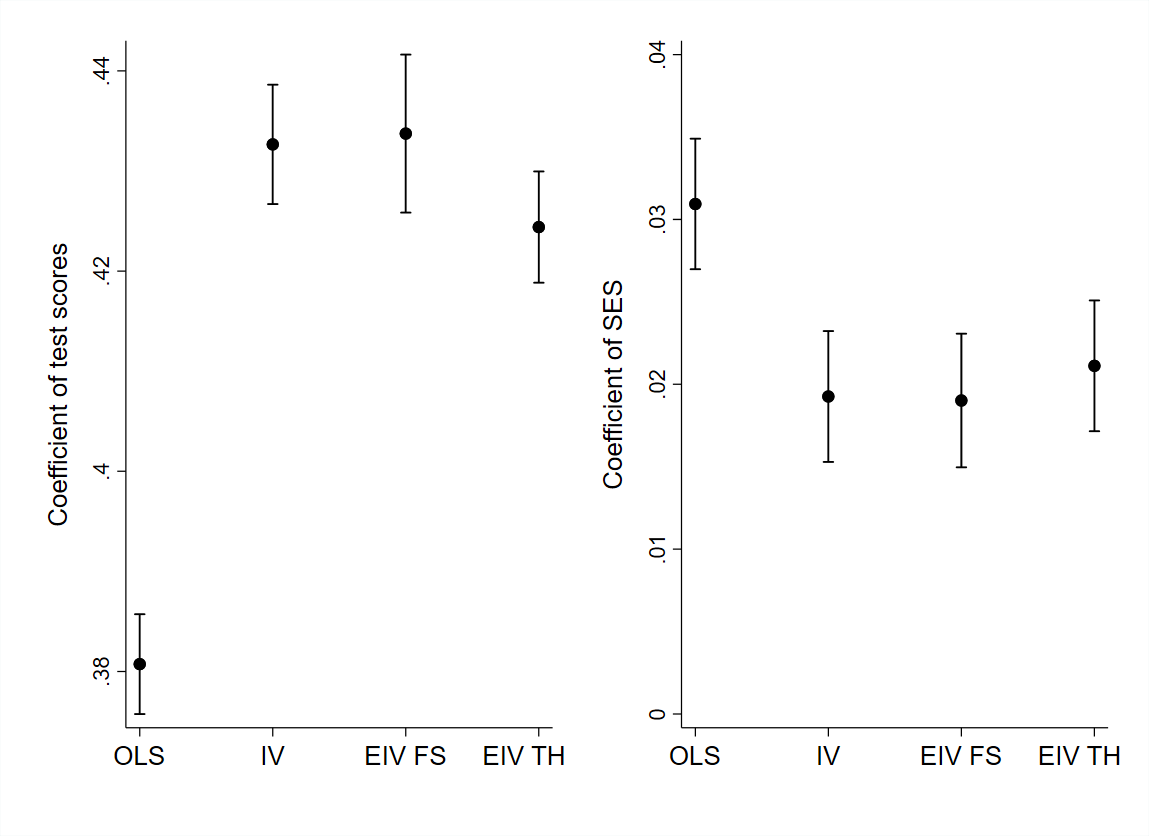}}\par 
\subfloat[Cohort 2020-21]{\label{c_hc}\includegraphics[width=.50\linewidth]{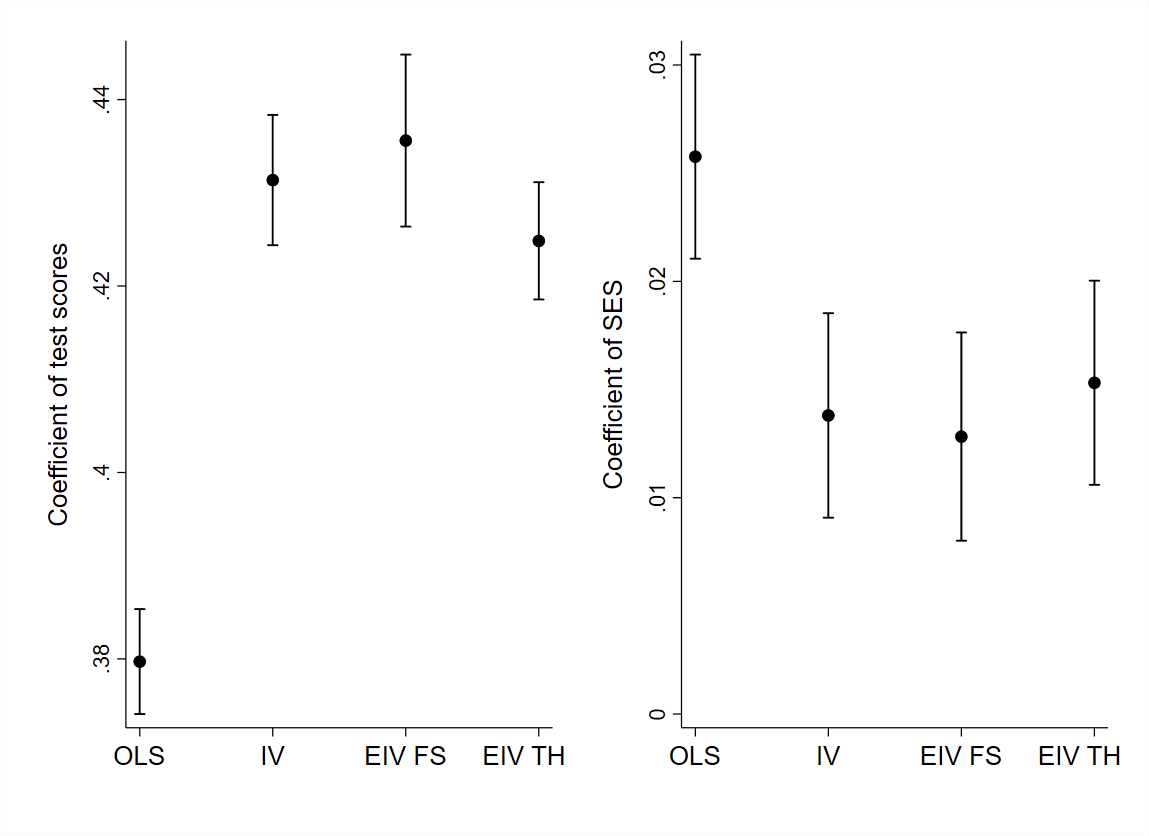}}
\subfloat[Cohort 2021-22]{\label{d_hc}\includegraphics[width=.50\linewidth]{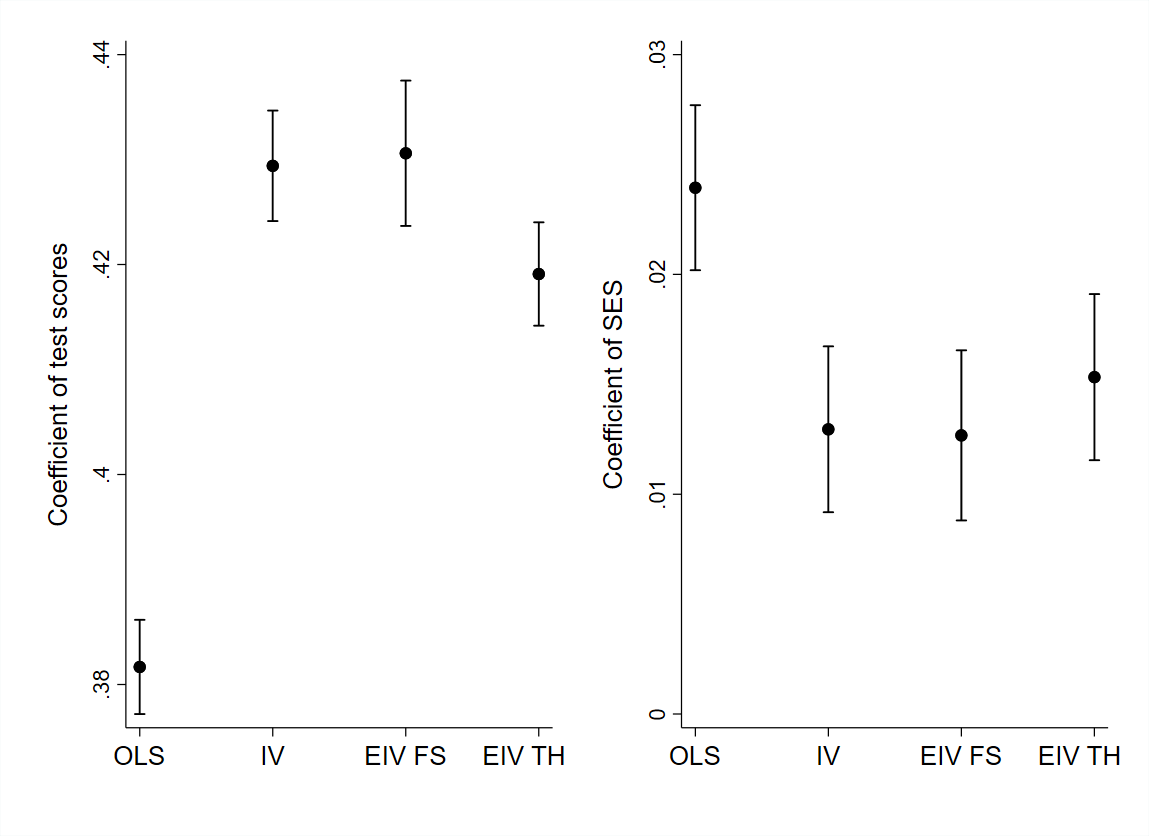}} 
\caption*{ \footnotesize
Notes: this figure shows the estimated coefficients for the OLS, IV, and EIV strategy, separately for each cohort (\Cref{fig:OLS_IV_EIV_cohorts}(a)-(d)). The lines are 95\% confidence intervals. See \Cref{fig:OLSIVEIV} for details on the estimation.  
}
\end{figure}

\begin{figure}
\centering
\caption{OLS, IV, and EIV estimates separately for immigrants and natives}
\label{fig:OLS_IV_EIV_imm}
\subfloat[Immigrants]{\label{a_hi}\includegraphics[width=.50\linewidth]{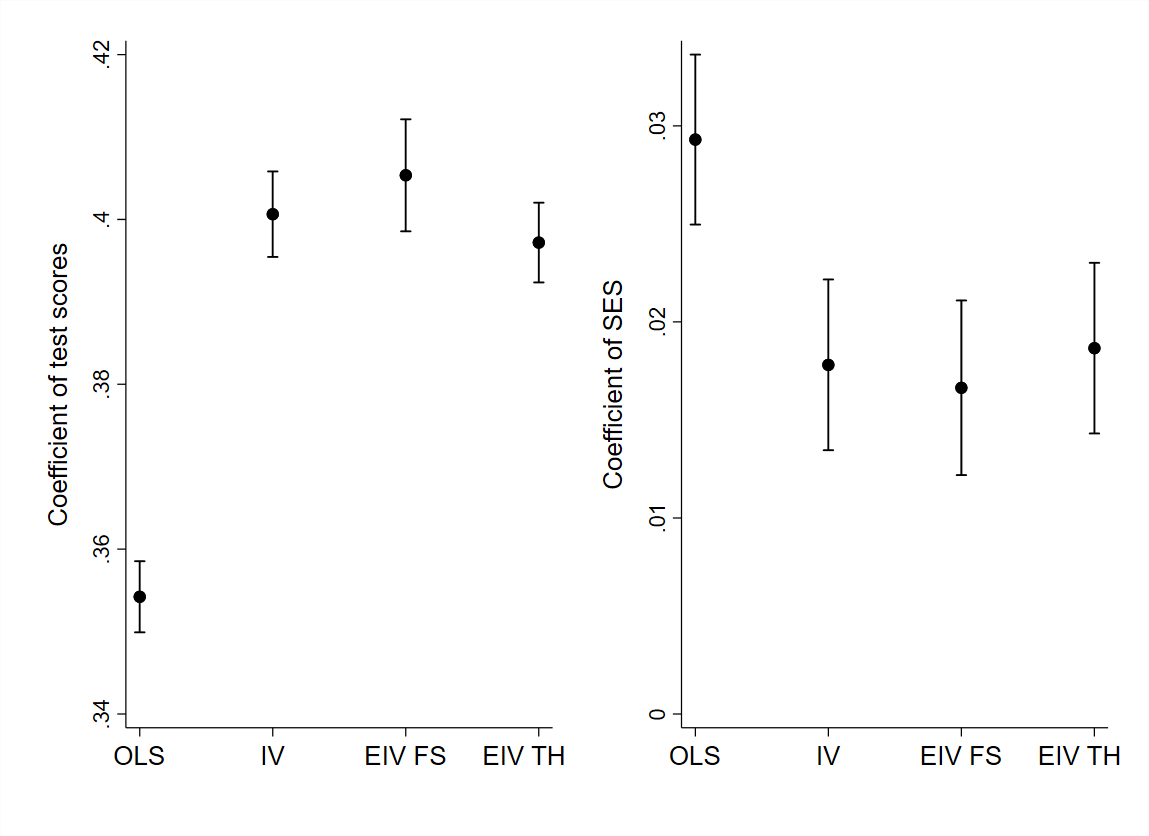}}
\subfloat[Natives]{\label{b_hi}\includegraphics[width=.50\linewidth]{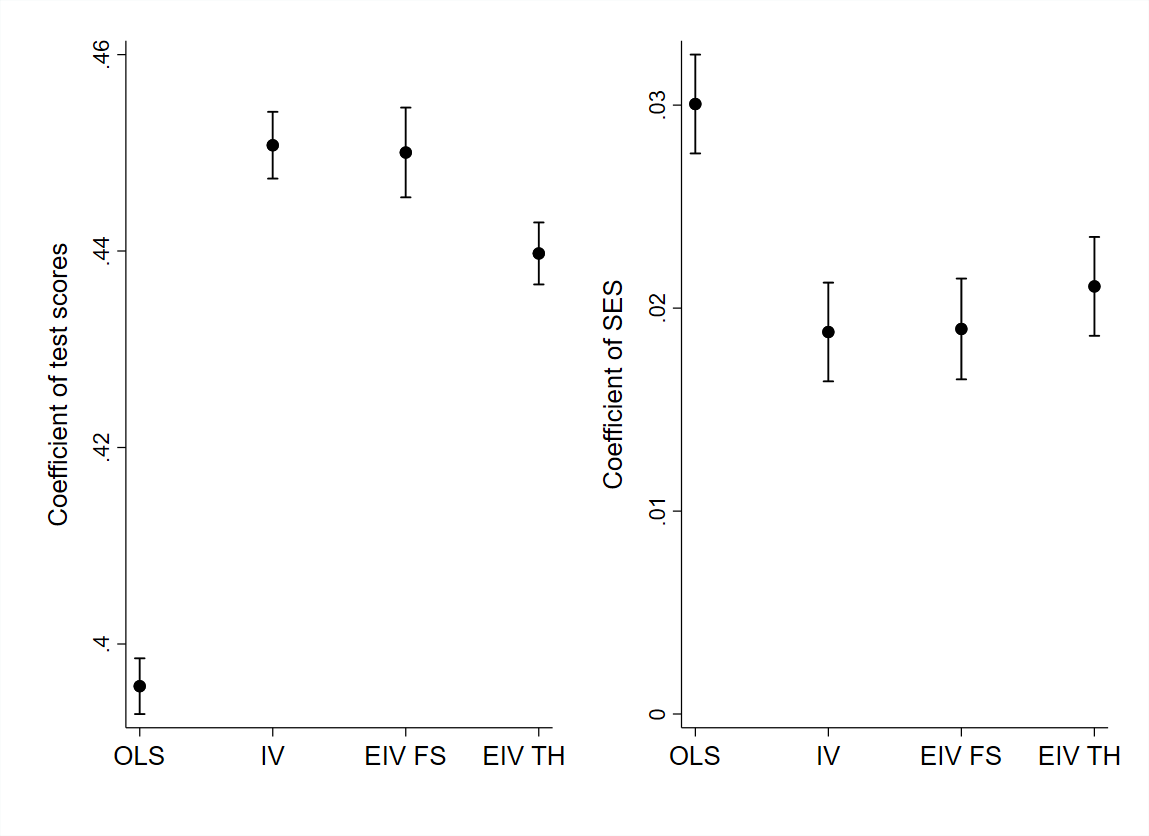}}
\caption*{ \footnotesize
Notes: this figure shows the estimated coefficients for the OLS, IV, and EIV strategy, separately for immigrants (\Cref{fig:OLS_IV_EIV_imm}(a)) and natives (\Cref{fig:OLS_IV_EIV_imm}(b)). The lines are 95\% confidence intervals. See \Cref{fig:OLSIVEIV} for details on the estimation.  
}
\end{figure}

\end{document}